\documentclass[draftclsnofoot,onecolumn,11pt,journal]{IEEEtran}

\usepackage{times}
\usepackage{epsfig}
\usepackage{graphicx}
\usepackage{amsmath}
\usepackage{amssymb}
\usepackage{amsthm}
\usepackage{enumerate}
\usepackage{indentfirst}
\usepackage{subfigure}
\usepackage{algorithmic}
\usepackage{algorithm}
\usepackage{url}
\usepackage{booktabs}
\usepackage{multirow}
\usepackage{longtable}
\usepackage{cite}

\theoremstyle{remark}
\newtheorem{theorem}{Theorem}

\newtheorem{lemma}{Lemma}
\newtheorem{corollary}{Corollary}

\long\def\symbolfootnote[#1]#2{\begingroup\def\thefootnote{\fnsymbol{footnote}}
\footnote[#1]{#2}\endgroup}

\ifCLASSINFOpdf
\else
\fi
\begin{document}
%
\title{An Analytical Framework for Heterogeneous Partial Feedback Design in Heterogeneous Multicell OFDMA Networks}
%
%
%

\author{Yichao Huang, \IEEEmembership{Member, IEEE}, and Bhaskar D. Rao, \IEEEmembership{Fellow,
IEEE}
\thanks{Copyright (c) 2012 IEEE. Personal use of this material is permitted. However, permission to use this material for any other purposes must be obtained from the IEEE by sending a request to pubs-permissions@ieee.org.}
\thanks{This research was supported by Ericsson endowed chair funds, the Center for Wireless Communications, UC Discovery grant com09R-156561 and NSF
grant CCF-1115645. The material in this paper was presented in part at the IEEE International Conference on Communications (ICC), Ottawa,
Canada, June 2012.}
\thanks{The authors are with Department of Electrical and Computer Engineering, University of California, San Diego, La Jolla, CA
92093-0407, USA (e-mail: yih006@ucsd.edu; brao@ece.ucsd.edu).}}

%
%

\markboth{To Appear in IEEE Transactions on Signal Processing}%
{To Appear in IEEE Transactions on Signal Processing}
%



\maketitle

\begin{abstract}
The inherent heterogeneous structure resulting from user densities and large scale channel effects motivates heterogeneous partial feedback
design in heterogeneous networks. In such emerging networks, a distributed scheduling policy which enjoys multiuser diversity as well as
maintains fairness among users is favored for individual user rate enhancement and guarantees. For a system employing the cumulative
distribution function based scheduling, which satisfies the two above mentioned desired features, we develop an analytical framework to
investigate heterogeneous partial feedback in a general OFDMA-based heterogeneous multicell employing the best-M partial feedback strategy.
Exact sum rate analysis is first carried out and closed form expressions are obtained by a novel decomposition of the probability density
function of the selected user's signal-to-interference-plus-noise ratio. To draw further insight, we perform asymptotic analysis using extreme
value theory to examine the effect of partial feedback on the randomness of multiuser diversity, show the asymptotic optimality of best-1
feedback, and derive an asymptotic approximation for the sum rate in order to determine the minimum required partial feedback.
\end{abstract}

\begin{IEEEkeywords}
Heterogeneous feedback, heterogeneous networks, multicell, OFDMA, partial feedback, multiuser diversity, extreme value theory
\end{IEEEkeywords}

%
\IEEEpeerreviewmaketitle

\section{Introduction}\label{introduction}
%
%
%
%

The growing dependence of users on wireless services will require wireless systems to become ubiquitous and offer seamless support. The demands
of video and other high data rate applications have placed increasing requirements on networks to support high data rate services in a cost
effective manner leading to heterogeneous networks. With the advent of OFDMA-based heterogeneous networks \cite{madan10, damnjanovic11} which
incorporate lower power pico \cite{damnjanovic12, lopez12}, femto base stations \cite{andrews12, barbieri12}, and fixed relays \cite{yang09,
sanguinetti12} to coexist with the traditional macrocell, the spectrum reuse has grown aggressively to full usage pattern across different tiers
\cite{dhillon11} of the heterogeneous structure. One challenging feature of heterogeneous networks is the self-created ``cell edges" within the
macrocell, which requires advanced techniques both in theory and in practice to model \cite{win09}, manage \cite{boudreau09}, and even make use
of the cross-tier intercell interference. Among the recent approaches that have been developed such as subcarrier allocation and power control
to carefully adapt the system resource in a centralized way \cite{ksairi10, son11, zhang11}, and utilization of spatial domain for cooperative
multicell processing to cancel, coordinate, and align the interference \cite{gesbert10, sawahashi10, dahrouj10, bhagavatula11, cadambe08,
suh11}, a significant impediment is the NP-hardness of the problem \cite{luo08}, the limited resource constraints, as well as the need for
extensive backhaul capability. These challenges favor the development of distributed solutions. All the aforementioned techniques in the
downlink assume the availability of channel state information at the transmitter (CSIT) via feedback\footnote[1]{For the purpose of scheduling
or optimization in a multicell network, the feedback is often needed in a frequency division duplex (FDD) system, or in a time division duplex
(TDD) system when the channel reciprocity can not be observed due to duplexing time delay.} \cite{love08} to adapt the network transmission
strategy to the varying wireless environment. With the rapidly growing number of wireless users, the amount of feedback for the OFDMA-based
networks may become prohibitive which motivates the design of efficient feedback schemes without significantly degrading system performance.

In addition to the usual challenges, there are two new issues that arise when investigating partial feedback in heterogeneous networks. Firstly,
due to the different locations of the users and different ranges of transmit powers, users' large scale channel effects are highly asymmetric.
Therefore, it is equally important to guarantee fairness among users as well as leveraging multiuser diversity \cite{knopp95, viswanath02} in an
opportunistic scheduling framework. Secondly, since the number of users or user densities are diverse in a heterogeneous network, it would be
beneficial to adapt the feedback and require less feedback when a serving base station has more users associated with it. We refer to this
methodology as heterogeneous partial feedback\footnote[2]{Note that achieving adaptive feedback according to the number of users, the users'
channel condition, and users' data rate requirements etc is an important issue in practical systems such as LTE \cite{dahlman11, sesia11}.} and
we aim to provide an analytical framework to identify its benefits under a fair and distributed opportunistic scheduling policy to fulfill the
vision of location awareness \cite{huang10c} and situational awareness.

The first step towards examining the aforementioned issues is developing an opportunistic scheduling policy, which exploits multiuser diversity
and also preserves scheduling fairness among heterogeneous users. Traditional scheduling policies such as round robin strategy \cite{hahne91}
and greedy strategy \cite{knopp95} are easy to implement, yet only achieve one of the desired features. In this paper, we consider a system that
employs the cumulative distribution function (CDF) based scheduling policy \cite{park05, patil09}. According to the basic CDF-based scheduling
strategy, an user is selected whose rate is high enough, but least probable to grow higher. Therefore, this scheduling strategy possesses
properties similar to the proportional fair scheduler \cite{kelly97, jalali00, viswanath02, choi07, caire07}, and additionally enables a user's
rate to be independent of the statistics of other users. Herein, the CDF-based scheduling policy is analyzed for a general OFDMA downlink in a
multicell environment to examine heterogeneous partial feedback design. Currently,  in an OFDMA-based system which groups subcarriers into
resource blocks \cite{zhu09, chen08, kuhne08} to form the basic scheduling and feedback unit, two partial feedback strategies are appealing: the
thresholding-based partial feedback \cite{sanayei07, hassel07, pugh10} and the best-M partial feedback \cite{jung07, ko07, choi08, leinonen09,
donthi11, hur11}. The latter strategy, which is considered in practical systems such as LTE \cite{dahlman11, sesia11}, requires the users to
order and convey the M best channels. Herein, we employ the best-M partial feedback strategy for further analysis. Intuitively, M would be
chosen to be small when the user density in a given cell is large, which motivates the utilization of heterogeneous feedback resource across
different cells in the heterogeneous networks.

Rigorous development of the analytical framework requires investigation of the interplay between the scheduling policy, partial feedback, and
the statistical property of the user's signal-to-interference-plus-noise ratio ($\mathsf{SINR}$). There are limited results available in the
literature on this topic and the available results analyzing the best-M partial feedback are for the single cell scenario without intercell
interference \cite{jung07, ko07, choi08, leinonen09, donthi11, hur11}. A detailed treatment of the best-M partial feedback is provided in
\cite{hur11} and a convenient polynomial form for the CDF of selected user's $\mathsf{SINR}$ is presented for analytical evaluation. Though it
is derived for the single cell scenario, it forms the building block for our heterogeneous network analysis which takes into account the
cross-tier intercell interference. In this paper, the analytical framework is treated first from the perspective of exact system performance. We
derive the closed form expression for the sum rate with the CDF-based scheduling policy and best-M partial feedback strategy. One key technique
developed and utilized is the decomposition of the probability density function (PDF) of the selected user's $\mathsf{SINR}$, which is amenable
for further integration needed to determine system performance. The derived closed form results are directly applicable to further system
evaluation.

In order to gain additional insight, we investigate the system performance from the asymptotic perspective utilizing extreme value theory
\cite{galambos78, david03} when the number of users in a given cell grows large \cite{sharif05, song06, yoo07, gesbert11, tajer11, bang12}.
Different from the special case of full feedback, examining the general best-M partial feedback incurs additional difficulties due to the
two-stage maximization resulting from partial feedback and scheduling policy, with the first stage maximization being performed at the user side
to provide selective feedback and the second stage maximization being performed at the scheduler side for user selection. Herein, we analyze the
tail behavior of the selected user's $\mathsf{SINR}$ and establish the type of convergence in order to examine the effect of partial feedback on
the randomness of multiuser diversity, show the optimality of best-1 feedback in the asymptotic sense, and more importantly, provide the
asymptotic approximation for the sum rate of the general best-M partial feedback. The established asymptotic results further help in
analytically tracking and determining the minimum required partial feedback.

To summarize, the contributions of this paper are threefold: a conceptual framework for situational-aware heterogeneous partial feedback design
in an OFDMA-based heterogeneous multicell network, a thorough analysis and derivation of closed form results for the sum rate, and a detailed
investigation of the partial feedback based on extreme value theory. All these contributions foster the understanding of heterogeneous feedback
design in future systems. Furthermore, the analytical tools developed promise to have broad applicability and can be applied to many related
problems. The remainder of the paper is organized as follows. The system model is provided in Section \ref{system}. The general treatment
without specific channel models is examined in Section \ref{scheduling}. By assuming standard channel models, Section \ref{exact} carries out
exact performance analysis, and Section \ref{asymptotic} presents asymptotic analysis. Numerical results are provided in Section
\ref{numerical}, and Section \ref{conclusion} concludes the paper.

\section{System Model}\label{system}
\begin{figure}[t]
\centering
    \includegraphics[width=0.6\linewidth]{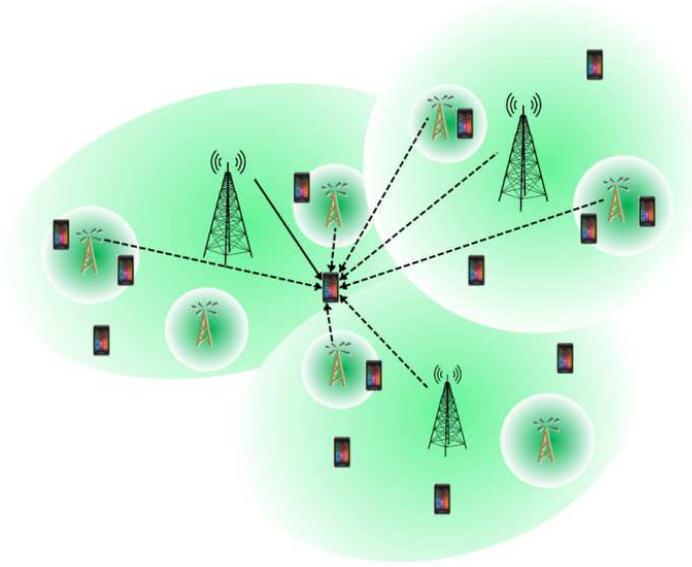}
\caption{Illustration of a generic OFDMA-based multicell heterogeneous cellular networks. Each cell has users associated with it. One selected
user for transmission in one resource block is shown: solid line (desired signal); dashed line (potential intercell interference).}
\label{fig_1}
\end{figure}

We consider the downlink of an OFDMA-based heterogeneous network. The model assumed is generic and sufficiently general to be applicable to a
multitier multicell network\footnote[3]{The special case with one picocell inside a macrocell under symmetric large scale channel effects is
studied in our recent work \cite{huang11letter}.}, e.g., see Fig. \ref{fig_1} for illustration. The system consists of $N$ resource blocks, with
one resource block as the basic feedback and scheduling unit. Full spectrum reuse is assumed and it is also assumed that there is no advanced
technique employed to suppress interference such as multiuser detection at the receiver side. The process of cell association is assumed to be
performed in advance. Without loss of generality, one base station $B_0$ from the base station set $\mathcal{B}$ and its associated users
$\mathcal{K}_0=\{1,\ldots,k,\ldots,K_0\}$ with $|\mathcal{K}_0|=K_0$ are considered.

The received signal $y_{k,n}^{(0)}$ of user $k$ at resource block $n$ is represented by
\begin{equation}\label{system_eq_1}
y_{k,n}^{(0)}=\sqrt{G_k^{(0)}}H_{k,n}^{(0)}s_{k,n}^{(0)}+\sum_{b=1}^{J_k}\sqrt{G_k^{(b)}}H_{k,n}^{(b)}s_{n}^{(b)}+v_{k,n}^{(0)},\quad k\in
\mathcal{K}_0,
\end{equation}
where the superscript indicates the base station, i.e., the serving cell and the interfering cells; $J_k$ denotes the number of effective
interfering cells for user $k$, with the influence of other interfering cells, namely the residual interference, included in the additive white
noise $v_{k,n}^{(0)}$ distributed with $\mathcal{CN}(0,\sigma_k^2).$ $s_{k,n}^{(0)}$ and $s_n^{(b)}$ are the transmitted symbols by the serving
cell and the interfering cell $B_b$ with $\mathbb{E}\left[|s_{k,n}^{(0)}|^2\right]=p^{(0)}$ and $\mathbb{E}\left[|s_n^{(b)}|^2\right]=p^{(b)}$.
$H_{k,n}^{(0)}$ and $H_{k,n}^{(b)},$ which are assumed to be independent across users and resource blocks\footnote[4]{This assumption
corresponds to the frequency domain block fading channel model \cite{chen08, donthi11, hur11} due to its simplicity and capability to provide a
good approximation to actual physical channels.}, denote the small scale frequency domain channel transfer function between the serving cell and
user $k$ at resource block $n$, and between the interfering cell $B_b$ and user $k$ at resource block $n$, respectively. $G_k^{(0)}$ and
$G_k^{(b)}$ represent the large scale channel gain between the serving cell and user $k$, and between the interfering cell $B_b$ and user $k$
respectively. Based on the aforementioned assumption, the $\mathsf{SINR}$ of user $k$ at resource block $n$ can be written as
\begin{equation}\label{system_eq_2}
\mathsf{SINR}_{k,n}^{(0)}=\frac{G_k^{(0)}p^{(0)}|H_{k,n}^{(0)}|^2}{\sum_{b=1}^{J_k}G_k^{(b)}p^{(b)}|H_{k,n}^{(b)}|^2+\sigma_k^2}
=\frac{\rho_k^{(0)}|H_{k,n}^{(0)}|^2}{\sum_{b=1}^{J_k}\rho_k^{(b)}|H_{k,n}^{(b)}|^2+1},
\end{equation}
where $\rho_k^{(0)}\triangleq \frac{G_k^{(0)}p^{(0)}}{\sigma_k^2}$, $\rho_k^{(b)}\triangleq \frac{G_k^{(b)}p^{(b)}}{\sigma_k^2}$. The
$\mathsf{SINR}$ is the channel quality information (CQI) that will be fed back and used for scheduling as discussed next.

\section{General Sum Rate Analysis with CDF Scheduling Policy and \\Best-M Partial Feedback}\label{scheduling}
This section is devoted to the analysis of the interplay between the scheduling policy and partial feedback for a general channel model (note
that no assumption on the distribution of the large and small scale channel gains has been made so far), with treatment of specific channel
model in Section \ref{exact}.

Let $Z_{k,n}^{(0)}$ represent $\mathsf{SINR}_{k,n}^{(0)}$ for notational simplicity and denote it as the CQI of user $k$ at resource block $n$
with CDF $F_{Z_{k}^{(0)}}$. The CDF does not depend on the resource block index $n$ because the $\mathsf{SINR}_{k,n}^{(0)}$'s are independent
and identically distributed (i.i.d.) across resource blocks $n$ for a given user $k$. Now consider the feedback procedure utilizing the best-M
partial feedback strategy. According to the best-M partial feedback strategy, users measure CQI for each resource block at their receiver and
feed back the CQI values of the $M$ best resource blocks chosen from the total of $N$ values\footnote[5]{We assume the CQI is fed back without
feedback delay.}. More details on the best-M partial feedback approach can be found in \cite{choi08, leinonen09, donthi11, hur11, huang12c}.
This selective feedback procedure involves a maximization stage at each user. Because only a subset of the ordered CQI are fed back, from the
perspective of the scheduler (i.e., the serving base station $B_0$), if it receives feedback on a certain resource block from a user, it is
likely to be any one of the CQI from the ordered subset. We now aim to find the CDF of the CQI seen at the scheduler side as a consequence of
partial feedback. Denote $Y_{k,n,M}^{(0)}$ as the received CQI at the scheduler for user $k$ at resource block $n$ under best-M partial
feedback, which is the outcome of the user side maximization. Let $F_{Y_{k,M}^{(0)}}$ be its CDF, with resource block index $n$ dropped due to
the i.i.d. property across resource blocks for a given user. It is easy to see that for the full feedback case, i.e., $M=N$,
$F_{Y_{k,N}^{(0)}}=F_{Z_k^{(0)}}$, and for the best-1 feedback case, i.e., $M=1$, $F_{Y_{k,1}^{(0)}}=(F_{Z_k^{(0)}})^N$. Utilizing the results
in \cite{hur11}, the CDF for the general best-M feedback case can be expressed as
\begin{equation}\label{system_eq_3}
F_{Y_{k,M}^{(0)}}(x)=\sum_{m=0}^{M-1}\xi_1(N,M,m)(F_{Z_k^{(0)}}(x))^{N-m},
\end{equation}
where $\xi_1(N,M,m)=\sum_{i=m}^{M-1}\frac{M-i}{M}{N\choose i}{i\choose m}\left(-1\right)^{i-m}$.

After the scheduler receives feedback from its serving users, it is ready to perform scheduling. It is clear that for the single cell scenario
without intercell interference, the scheduling policy is easier to implement and analyze. For instance, in the single cell scenario with
homogeneous users, namely same large scale effects, the greedy scheduler or the max-$\mathsf{SNR}$ scheduler makes full use of multiuser
diversity as well as guarantees fairness due to the same statistics of the user's CQI. In the single cell scenario with heterogeneous users,
i.e., different large scale effects \cite{hur11}, the normalized greedy scheduler which selects user according to their normalized CQI has the
same desired property. However, in the general multicell scenario with intercell interference, the $\mathsf{SINR}_{k,n}^{(0)}$'s are independent
but non-identically distributed (i.n.i.d.) across users. Therefore, the received CQI at the scheduler for different users $Y_{k,n,M}^{(0)}$ are
i.n.i.d. across users. In order to leverage multiuser diversity and guarantee fairness\footnote[6]{Note that the motivations as well as the
fairness for the proportional-fair (PF) scheduling policy and CDF-based scheduling policy are very different. The PF policy targets the system
utility as the definition of system fairness. The CDF-based policy targets the long-term user fairness and each user on average is equiprobable
to be scheduled. In the single cell case, it can be shown that these two scheduling policies have similar effects. However, in the general
multicell case, users' rates are coupled under the PF policy, and independent under the CDF-based policy. Analyzing and comparing these two
scheduling policies in the general multicell networks are left to our future work.}, we employ the CDF-based scheduling policy \cite{park05,
patil09}. According to the CDF-based scheduling policy, the scheduler will utilize the distribution of the received CQI for each user, i.e.,
$F_{Y_{k,M}^{(0)}}$. Herein, it is assumed that the scheduler perfectly knows the CDF\footnote[7]{This is the only system requirement to perform
CDF-based scheduling, and the CDF can be obtained by infrequent feedback from users and learned by the system. Methods to estimate the CDF can
be found in \cite{park05}.}, and it conducts the following transformation
\begin{equation}\label{system_eq_4}
\tilde{Y}_{k,n,M}^{(0)}=F_{Y_{k,M}^{(0)}}(Y_{k,n,M}^{(0)}).
\end{equation}
The transformed random variable $\tilde{Y}_{k,n,M}^{(0)}$ is uniformly distributed over the range from $0$ to $1$ and
can be regarded as the virtual received CQI of user $k$ at resource block $n$.
The transformed random variables $\tilde{Y}_{k,n,M}^{(0)}$'s are i.i.d. across
users, which enables the maximization at the scheduler side to perform fair
scheduling. Denoting $k_n^*$ as the random variable representing the
selected user for transmission at resource block $n$, then
\begin{equation} \label{system_eq_5}
k_n^*=\arg\max_{k\in \mathcal{U}_{n,M}}\tilde{Y}_{k,n,M}^{(0)},
\end{equation}
where $\mathcal{U}_{n,M}$ denotes the set of users who convey feedback for resource block $n$. It can be easily seen that when $M=N$,
$\mathbb{P}(|\mathcal{U}_{n,N}|=K_0)=1$. For the general case when $1\leq M<N$, the probability mass function (PMF) of $|\mathcal{U}_{n,M}|$ can
be shown to be
\begin{equation} \label{system_eq_6}
\mathbb{P}(|\mathcal{U}_{n,M}|=\tau_0)={K_0\choose \tau_0}\left(\frac{M}{N}\right) ^{\tau_0}\left(1-\frac{M}{N}\right)^{K_0-\tau_0}, 0\leq
\tau_0\leq K_0.
\end{equation}
After the user $k_n^*$ is selected according to (\ref{system_eq_5}), the scheduler utilizes the corresponding $Y_{k_n^*,n,M}^{(0)}$ for rate
matching of the selected user. We denote the random variable $X_{n,M}^{(0)}$ as the selected user's CQI for resource block $n$, and use
the sum rate as the system performance metric. The sum rate
$C^{(0)}(M)$ for a given base station $B_0$ employing the CDF-based scheduling and best-M partial feedback is defined as follows
\begin{equation} \label{system_eq_14}
C^{(0)}(M)=\frac{1}{N}\sum_{n=1}^N\mathbb{E}\left[\log_2\left(1+X_{n,M}^{(0)}\right)\right].
\end{equation}
From the aforementioned analysis, the sum rate can be formulated, with appropriate conditioning, as\footnote[8]{In order to maintain full
frequency reuse for analytical tractability, it is assumed that if no user provides CQI for a certain resource block, then that resource block
would be in outage and would not contribute to the sum rate calculation.}
\begin{align}
C^{(0)}(M)&=\frac{1}{N}\sum_{n=1}^N\mathbb{E}_{k_n^*}\mathbb{E}_{|\mathcal{U}_{n,M}|}\left[\mathbb{E}_{X_{n,M}^{(0)}}\left[\log_2\left(1+X_{n,M}^{(0)}\right)\mid
|\mathcal{U}_{n,M}|\neq0\right]+\mathbb{E}_{X_{n,M}^{(0)}}\left[\log_2\left(1+X_{n,M}^{(0)}\right)\mid |\mathcal{U}_{n,M}|=0\right]\right]\notag\\
&=\frac{1}{N}\sum_{n=1}^N\mathbb{E}_{k_n^*}\mathbb{E}_{|\mathcal{U}_{n,M}|}\left[\int_0^1\log_2\left(1+F_{Y_{k_n^*,M}^{(0)}}^{-1}(x)\right)
dx^{\tau_0}\mid |\mathcal{U}_{n,M}|\neq0\right]\notag\\
\label{system_eq_7}&\mathop{=}\limits^{(a)}\mathbb{E}_{k^*}\mathbb{E}_{|\mathcal{U}_{M}|}\left[\int_0^{\infty}\log_2(1+t)
d(F_{Y_{k^*,M}^{(0)}}(t))^{\tau_0}\mid |\mathcal{U}_{M}|\neq0\right],
\end{align}
where (a) follows from the identical distributed property across resource blocks and the change of variable $x=F_{Y_{k^*,M}^{(0)}}(t)$. The
conditional statistical property of $X_{n,M}^{(0)}$ conditioned on the selected user $k_n^*$ and the set of users who have conveyed feedback
$\mathcal{U}_{n,M}$ can be expressed as
\begin{equation} \label{system_eq_8}
F_{X_{M}^{(0)}\mid k^*=k,|\mathcal{U}_{M}|=\tau_0}(x)=(F_{Y_{k,M}^{(0)}}(x))^{\tau_0}.
\end{equation}
Using (\ref{system_eq_3}), it can be expressed in the following power series expansion \cite{gradshteyn07, hur11}
\begin{equation} \label{system_eq_9}
F_{X_{M}^{(0)}\mid k^*=k,|\mathcal{U}_{M}|=\tau_0}(x)=\sum_{m=0}^{\tau_0(M-1)}\xi_2(N,M,\tau_0,m)(F_{Z_k^{(0)}}(x))^{N\tau_0-m},
\end{equation}
where
\begin{equation} \label{system_eq_10}
\xi_2(N,M,\tau_0,m)=\left\{
\begin{array}{l}
(\xi_1(N,M,0))^{\tau_0},\quad m=0\\ \frac{1}{m\xi_1(N,M,0)}\sum_{\ell=1}^{\min(m,M-1)}((\tau_0+1)\ell-m)\\
\quad\times\xi_1(N,M,\ell)\xi_2(N,M,\tau_0,m-\ell),\quad 1\leq m<\tau_0(M-1)\\ (\xi_1(N,M,M-1))^{\tau_0},\quad m=\tau_0(M-1).\\
\end{array} \right.
\end{equation}
Using (\ref{system_eq_6}) and (\ref{system_eq_9}), the sum rate (\ref{system_eq_7}) can be expressed in the following form
\begin{align}
C^{(0)}(M)&=\mathbb{E}_{k^*}\mathbb{E}_{|\mathcal{U}_{M}|}\left[\int_0^{\infty}\log_2(1+x)dF_{X_{M}^{(0)}\mid
k^*=k,|\mathcal{U}_{M}|=\tau_0}(x)\mid |\mathcal{U}_{M}|\neq0\right]\notag\\
\label{system_eq_11}&\mathop{=}\limits^{(a)}\frac{1}{K_0}\sum_{k=1}^{K_0}\sum_{\tau_0=1}^{K_0}{K_0\choose \tau_0}\left(\frac{M}{N}\right)
^{\tau_0}\left(1-\frac{M}{N}\right)^{K_0-\tau_0}\sum_{m=0}^{\tau_0(M-1)}\xi_2(N,M,\tau_0,m)\mathcal{G}_k(N\tau_0-m),
\end{align}
where (a) follows from the fair property of the CDF-based scheduling policy:
$\mathbb{P}\left(k^*=k,|\mathcal{U}_M|=\tau_0\right)=\frac{1}{K_0}\mathbb{P}\left(|\mathcal{U}_M|=\tau_0\right)$. The integration
$\mathcal{G}_k(\epsilon)$ for $\epsilon\in\mathbb{N}_+$ is defined as
\begin{equation} \label{system_eq_12}
\mathcal{G}_k(\epsilon)\triangleq \int_0^{\infty}\log_2(1+x)d(F_{Z_k^{(0)}}(x))^{\epsilon}.
\end{equation}
From (\ref{system_eq_11}), the individual user rate for user $k$ can be expressed as
\begin{equation} \label{system_eq_13}
C_k^{(0)}(M)=\frac{1}{K_0}\sum_{\tau_0=1}^{K_0}{K_0\choose \tau_0}\left(\frac{M}{N}\right)
^{\tau_0}\left(1-\frac{M}{N}\right)^{K_0-\tau_0}\sum_{m=0}^{\tau_0(M-1)}\xi_2(N,M,\tau_0,m)\mathcal{G}_k(N\tau_0-m).
\end{equation}
For the special full feedback case, the sum rate becomes $C^{(0)}(N)=\frac{1}{K_0}\sum_{k=1}^{K_0}\mathcal{G}_k(K_0)$, and the individual user
rate for user $k$ becomes $C_k^{(0)}(N)=\frac{1}{K_0}\mathcal{G}_k(K_0)$.

\textit{Remark:} A few remarks are in order. Firstly, the effect of best-M partial feedback and the CDF-scheduling policy result in a two stage
maximization. The first stage maximization occurs at each user side to select the M best CQI for feedback. The second stage maximization is
conducted at the scheduler side by performing CDF-based transformation and user scheduling. Secondly, with the help of CDF-based scheduling,
each user feels as if the other users had the same CDF for scheduling competition \cite{park05}. In other words, each individual user's rate is
independent of other users. This important feature not only enables the distributed system to enjoy multiuser diversity, but also makes it
possible to consider or predict each user's rate by only considering its own CDF. Thirdly, users are equiprobable to be scheduled despite of
their heterogeneous channels (e.g., different statistics due to diverse propagation environments and interference levels), and so the scheduling
policy maintains fairness among users.

Up to now, we have obtained the general form of the sum rate and individual user rate with the help of $\mathcal{G}_k(\epsilon)$ without
assuming specific distributions on the channel models. In the next section, we derive the closed form expression for
$\mathcal{G}_k(\epsilon)$ with standard channel models.

\section{Exact Performance Analysis for Rayleigh Fading Channels}\label{exact}
In this section, we perform exact analysis to derive the closed form sum rate with standard Rayleigh fading
channel models. Section \ref{statistics} examines
the PDF and CDF of the $\mathsf{SINR}$ for each user and uses
them to derive a closed form expression for $\mathcal{G}_k(\epsilon)$ in Section
\ref{procedure}.

\subsection{The Statistics of CQI}\label{statistics}
In a practical system setting, the time scale for the large scale and small scale channel effects are much different. The variation of the small
scale channel gain $H$ occurs on the order of millisecond; whereas the large scale channel gain $G$ which may consist of path loss, antenna
gain, and shadowing, varies usually on the order of tens of seconds. Therefore, the large scale channel effect is assumed to be known in advance
by the system, through infrequent feedback or location awareness. The small scale channel effect is modeled as complex Gaussian distributed
random variables with zero mean and unit variance $\mathcal{CN}(0,1)$. From the definition of $\mathsf{SINR}$ in (\ref{system_eq_2}), it can be
seen that the numerator is a scaled $\chi^2(2)$ random variable (i.e., chi-square random variable with $2$ degrees of freedom), and the
denominator is a weighted sum of $\chi^2(2)$ random variables plus a constant. The following lemma provides the density function of
$\mathsf{SINR}_k^{(0)}$, namely $f_{Z_k^{(0)}}$.
\begin{lemma} \label{lemma_1}
The PDF of $Z_k^{(0)}$ can be expressed as
\begin{equation} \label{exact_eq_1}
f_{Z_k^{(0)}}(x)=\sum_{b=1}^{J_k}\varpi_k^{(b)}e^{-\frac{x}{\rho_k^{(0)}}}\left(\frac{1}{\rho_k^{(0)}+\rho_k^{(b)}x}+\frac{\rho_k^{(0)}\rho_k^{(b)}}{\left(\rho_k^{(0)}+\rho_k^{(b)}x\right)^2}\right)u(x),
\end{equation}
where $\varpi_k^{(b)}=\mathop{\prod}\limits_{\substack{i=1\\i\neq b}}\frac{\rho_k^{(b)}}{\rho_k^{(b)}-\rho_k^{(i)}}$, and $u(\cdot)$ is the
Heaviside step function.
\end{lemma}
\begin{proof}
The proof is given in Appendix \ref{appenA}.
\end{proof}
From Lemma \ref{lemma_1}, the CDF of $Z_k^{(0)}$, namely $F_{Z_k^{(0)}}$ can be computed as
\begin{align}
F_{Z_k^{(0)}}(x)&=\int_0^x\sum_{b=1}^{J_k}\varpi_k^{(b)}e^{-\frac{x}{\rho_k^{(0)}}}\left(\frac{1}{\rho_k^{(0)}+\rho_k^{(b)}y}+\frac{\rho_k^{(0)}\rho_k^{(b)}}{\left(\rho_k^{(0)}+\rho_k^{(b)}y\right)^2}\right)dy\notag\\
\label{exact_eq_2}&=\left(1-\sum_{b=1}^{J_k}\frac{\varpi_k^{(b)}e^{-\frac{x}{\rho_k^{(0)}}}\rho_k^{(0)}}{\rho_k^{(0)}+\rho_k^{(b)}x}\right)u(x).
\end{align}

\subsection{Procedures to Compute $\mathcal{G}_k(\epsilon)$}\label{procedure}
Now we consider the computation of $\mathcal{G}_k(\epsilon)=\int_0^{\infty}\log_2(1+x)d(F_{Z_k^{(0)}}(x))^{\epsilon}$, which will be carried out
in three steps. Step $1$ provides a suitable PDF decomposition of $d(F_{Z_k^{(0)}}(x))^{\epsilon}$ by examining the expression for the PDF. In
Step $2$, the decomposed PDF is further expanded for integration. Finally, Step $3$ employs the outcome of Step $1$ and $2$ to derive the closed
form expression for $\mathcal{G}_k(\epsilon)$ by standard integration techniques. The details are presented next.

\textit{Step $1$:} We are interested in the exact formulation of $d(F_{Z_k^{(0)}}(x))^{\epsilon}$, where the exponent $\epsilon\in
\mathbb{N}_+$. In the following lemma, an amenable decomposition is proposed for the statistical form of $d(F_{Z_k^{(0)}}(x))^{\epsilon}$.
\begin{lemma} \label{lemma_2}
The PDF $d(F_{Z_k^{(0)}}(x))^{\epsilon}$ with $\epsilon\in \mathbb{N}_+$ can be decomposed as
\begin{equation}\label{exact_eq_3}
d(F_{Z_k^{(0)}}(x))^{\epsilon} = \epsilon\sum_{\ell=0}^{\epsilon-1}{\epsilon-1\choose \ell}\frac{(-1)^{\ell}}{\ell+1}
\;d\left(1-e^{-\frac{(\ell+1)x}{\rho_k^{(0)}}}\left(\sum_{b=1}^{J_k}\varpi_k^{(b)}\frac{\rho_k^{(0)}}{\rho_k^{(b)}}\frac{1}{x+\frac{\rho_k^{(0)}}{\rho_k^{(b)}}}\right)^{\ell+1}\right).
\end{equation}
\end{lemma}
\begin{proof}
The proof is given in Appendix \ref{appenA}.
\end{proof}

\textit{Step $2$:} Even though the complicated form of $d(F_{Z_k^{(0)}}(x))^{\epsilon}$ is decomposed into (\ref{exact_eq_3}), its formulation
still prevents direct integration. The following lemma provides an expanded form for one of the terms to facilitate further integration.
\begin{lemma} \label{lemma_3}
\begin{equation}\label{exact_eq_4}
\left(\sum_{b=1}^{J_k}\varpi_k^{(b)}\frac{\rho_k^{(0)}}{\rho_k^{(b)}}\frac{1}{x+\frac{\rho_k^{(0)}}{\rho_k^{(b)}}}\right)^{\ell+1}=\sum_{j_1+\cdots+j_{J_k}=\ell+1}{\ell+1
\choose
j_1,\ldots,j_{J_k}}\sum_{b=1}^{J_k}\sum_{i=0}^{j_b}\frac{\psi_{k,i}^{(b)}\mathop{\prod}\limits_{b=1}^{J_k}\left(\frac{\varpi_k^{(b)}\rho_k^{(0)}}{\rho_k^{(b)}}\right)^{j_b}}{\left(x+\frac{\rho_k^{(0)}}{\rho_k^{(b)}}\right)^i},
\end{equation}
where
\begin{equation}\label{exact_eq_5}
\psi_{k,i}^{(b)}=\left\{
\begin{array}{l}
0,\quad i=0\\
\frac{1}{(j_b-i)!}\frac{d^{j_b-i}}{dx^{j_b-i}}\left[\left(x+\frac{\rho_k^{(0)}}{\rho_k^{(b)}}\right)^{j_b}\mathop{\prod}\limits_{b=1}^{J_k}\left(x+\frac{\rho_k^{(0)}}{\rho_k^{(b)}}\right)^{-j_b}\right]\Bigg|_{x=-\frac{\rho_k^{(0)}}{\rho_k^{(b)}}},\quad i\geq 1.\\
\end{array} \right.
\end{equation}
\end{lemma}
\begin{proof}
The proof is given in Appendix \ref{appenA}.
\end{proof}
For illustration purpose, the formulation of Lemma \ref{lemma_3} is discussed and provided for the special $J_k=2$ case in Appendix
\ref{appenA}.

\textit{Step $3$:} The following theorem completes the final step by utilizing the outcomes of the above two steps to derive the closed form
expression for $\mathcal{G}_k(\epsilon)$.
\begin{theorem} \label{theorem_1}
$\mathcal{G}_k(\epsilon)$ can be computed as
\begin{align}
\mathcal{G}_k(\epsilon)&=\epsilon\sum_{\ell=0}^{\epsilon-1}{\epsilon-1\choose
\ell}\frac{(-1)^{\ell}}{\ell+1}\sum_{j_1+\cdots+j_{J_k}=\ell+1}{\ell+1 \choose
j_1,\ldots,j_{J_k}}\sum_{b=1}^{J_k}\sum_{i=0}^{j_b}\psi_{k,i}^{(b)}\notag\\
\label{exact_eq_6}&\quad\quad\times\mathop{\prod}\limits_{b=1}^{J_k}\left(\frac{\varpi_k^{(b)}\rho_k^{(0)}}{\rho_k^{(b)}}\right)^{j_b}\mathcal{I}_1\left(\frac{(\ell+1)}{\rho_k^{(0)}},
\frac{\rho_k^{(0)}}{\rho_k^{(b)}}, i\right),
\end{align}
where $\mathcal{I}_1(\alpha,\beta,\gamma)\triangleq \int_0^{\infty}\frac{e^{-\alpha x}}{(1+x)(\beta+x)^{\gamma}} dx$ whose closed form
expression is presented in Appendix \ref{appenA}.
\end{theorem}
\begin{proof}
The proof is given in Appendix \ref{appenA}.
\end{proof}
The three-step procedure yields the closed form expression for $\mathcal{G}_k(\epsilon)$, which can be substituted into (\ref{system_eq_11}) and
(\ref{system_eq_13}) to compute the closed form sum rate and individual user rate. The exact closed form expressions only involves finite sums
and factorials making it computationally tractable and useful for system evaluation. In the following, the treatment of two simplified special
cases are provided: the one-dominant interference limited case and the noise limited case.

\textit{One-Dominant Interference Limited Case:} This case approximates the scenario when there is one dominant interferer. Without loss of
generality, assume $\rho_k^{(1)}\gg \rho_k^{(b)}$ for $b\neq1$ and the effect of noise is omitted.
Then the $\mathsf{SINR}$ can be approximated
as $\mathsf{SINR}_{k,n}^{(0)}\simeq\mathsf{SIR}_{k,n}^{(0)}=\frac{\rho_k^{(0)}|H_{k,n}^{(0)}|^2}{\rho_k^{(1)}|H_{k,n}^{(1)}|^2}$, which is the F
distributed random variable. The CDF of the CQI can be written as
\begin{equation}\label{exact_eq_7}
F_{Z_k^{(0)}}(x)=\left(1-\frac{\rho_k^{(0)}}{\rho_k^{(1)}x+\rho_k^{(0)}}\right)u(x).
\end{equation}
In this case, the computation of $\mathcal{G}_k(\epsilon)$ can be reduced to
\begin{equation}\label{exact_eq_8}
\mathcal{G}_k(\epsilon)\mathop{=}\limits^{(a)}
\frac{\epsilon}{\ln2}\frac{\rho_k^{(0)}}{\rho_k^{(1)}}\sum_{\ell=0}^{\epsilon-1}{\epsilon-1\choose
\ell}\frac{(-1)^{\ell}}{\ell+1}\mathrm{Beta}(1,\ell+1){}_2F_1\left(1,1;\ell+2;1-\frac{\rho_k^{(0)}}{\rho_k^{(1)}}\right),
\end{equation}
where (a) follows from \cite[3.197.5]{gradshteyn07}, $\mathrm{Beta}(x,y)=\int_0^1 t^{x-1}(1-t)^{y-1}dt$ is the Beta function
\cite[8.38]{gradshteyn07} and ${}_2F_1(\cdot,\cdot;\cdot;\cdot)$ is the Gaussian hypergeometric function \cite{abramowitz72}.

\textit{Noise Limited Case:} This case approximates the scenario when the impact of intercell interference is negligible, i.e., $\rho_k^{(b)}\ll
1$. The $\mathsf{SINR}$ can be approximated as $\mathsf{SINR}_{k,n}^{(0)}\simeq\mathsf{SNR}_{k,n}^{(0)}=\rho_k^{(0)}|H_{k,n}^{(0)}|^2$, which is
the $\chi^2(2)$ distributed random variable. The CDF of the CQI can be written as
\begin{equation}\label{exact_eq_9}
F_{Z_k^{(0)}}(x)=\left(1-e^{-\frac{x}{\rho_k^{(0)}}}\right)u(x).
\end{equation}
In this case, the computation of $\mathcal{G}_k(\epsilon)$ can be reduced to
\begin{equation}\label{exact_eq_10}
\mathcal{G}_k(\epsilon)\mathop{=}\limits^{(a)} \frac{\epsilon}{\ln2}\sum_{\ell=0}^{\epsilon-1}{\epsilon-1\choose
\ell}\frac{(-1)^{\ell}}{\ell+1}e^{\frac{\ell+1}{\rho_k^{(0)}}}E_1\left(\frac{\ell+1}{\rho_k^{(0)}}\right),
\end{equation}
where (a) follows from \cite[4.337.2]{gradshteyn07} and $E_1(x)=\int_x^\infty \frac{e^{-t}}{t}dt$ is the exponential integral
function of the first order \cite{abramowitz72}. The noise limited case is equivalent to the single cell problem which has been addressed
extensively in \cite{choi08, leinonen09, hur11}.

\textit{Remark:} It can be easily seen that the CDF-based scheduling for the two simplified cases has the same effect as the ``normalized" CQI
based scheduling, which is normalized by $\frac{\rho_k^{(0)}}{\rho_k^{(1)}}$ for the one-dominant interference limited case and normalized by
$\rho_k^{(0)}$ for the noise limited case. The general CDF-based scheduling policy enables the general analysis for the multicell scenario, whose closed
form expressions have been obtained by the aforementioned procedures. The exact expression, though computable, is not easy to interpret and draw
insights. We now use asymptotic analysis to develop results that have the potential of providing further insights.

\section{Asymptotic Performance Analysis}\label{asymptotic}
This section is devoted to the asymptotic analysis when the associated users in a given cell grows large. Section \ref{type} proves the type of
convergence exhibited by the received CQI under best-M partial feedback. A brief summary on the different types
of convergence is provided in Appendix \ref{appenB} for easy reference.
In Section \ref{rate}, the asymptotic rate approximation is derived and is
employed to determine the minimum required partial feedback in Section \ref{determine}. The results are presented with the  proofs
relegated to the appendix.

\subsection{The Type of Convergence}\label{type}
The first step towards performing asymptotic analysis is examining the tail behavior of the received CQI at the scheduler side under partial
feedback, namely $Y_{k,M}^{(0)}$ for user $k$. In the full feedback case, $F_{Y_{k,N}^{(0)}}=F_{Z_k^{(0)}}$, which means the tail behavior of
the CQI at the user side $Z_k^{(0)}$ is equivalent to that of the $Y_{k,N}^{(0)}$. However, for the general best-M partial feedback, the
relationship between $Y_{k,M}^{(0)}$ and $Z_k^{(0)}$ is given by (\ref{system_eq_3}) and is recalled here for easy reference as
$F_{Y_{k,M}^{(0)}}(x)=\sum_{m=0}^{M-1}\xi_1(N,M,m)(F_{Z_k^{(0)}}(x))^{N-m}$. One natural question is concerning the relationship between the tail
behavior of $Y_{k,M}^{(0)}$ and $Z_k^{(0)}$, or formulated in a rigorous way: how to infer the type of convergence of $F_{Y_{k,M}^{(0)}}$ from
the type of convergence of $F_{Z_k^{(0)}}$ under the condition of best-M partial feedback? The following theorem addresses this  issue.
\begin{theorem} \label{theorem_2}
(\textit{Type of Convergence under Partial Feedback}) $F_{Y_{k,M}^{(0)}}$ has the same type of convergence property as $F_{Z_k^{(0)}}$ under the
best-M partial feedback strategy.
\end{theorem}
\begin{proof}
The proof is given in Appendix \ref{appenB}.
\end{proof}

Theorem \ref{theorem_2} states that the best-M partial feedback does not affect the type of convergence. In other words, once the type of
convergence for $F_{Z_k^{(0)}}$ is proven, the same property is established for $F_{Y_{k,M}^{(0)}}$. Note that so far no specific statistical
property has been assumed for $F_{Z_k^{(0)}}$. In the following, the statistical model expressed in (\ref{exact_eq_2}) will be utilized for further
analysis. The following corollary describes the tail behavior of $F_{Z_k^{(0)}}$.

\begin{corollary} \label{corollary_1}
For the general $\mathsf{SINR}$ case in the multicell scenario, $F_{Z_k^{(0)}}$ and $F_{Y_{k,M}^{(0)}}$ belong to the domain of attraction of
the Gumbel distribution \cite{galambos78}, i.e., $F_{Z_k^{(0)}}\in\mathcal{D}(G_3)$ and $F_{Y_{k,M}^{(0)}}\in\mathcal{D}(G_3)$.
\end{corollary}
\begin{proof}
The proof is given in Appendix \ref{appenB}.
\end{proof}

For completeness, the tail behavior of $F_{Z_k^{(0)}}$ for the special simplified one-dominant interference limited case and the noise limited
case is also provided in the following corollary.
\begin{corollary} \label{corollary_2}
For the one-dominant interference limited case, $F_{Z_k^{(0)}}$ and $F_{Y_{k,M}^{(0)}}$ belong to the domain of attraction of the Fr{\'e}chet
distribution \cite{galambos78}, i.e., $F_{Z_k^{(0)}}\in\mathcal{D}(G_1)$ and $F_{Y_{k,M}^{(0)}}\in\mathcal{D}(G_1)$. For the noise limited case,
$F_{Z_k^{(0)}}$ and $F_{Y_{k,M}^{(0)}}$ belong to the domain of attraction of Gumbel distribution \cite{galambos78}, i.e.,
$F_{Z_k^{(0)}}\in\mathcal{D}(G_3)$ and $F_{Y_{k,M}^{(0)}}\in\mathcal{D}(G_3)$.
\end{corollary}
\begin{proof}
The proof is given in Appendix \ref{appenB}.
\end{proof}
These established type of convergence results will be used to obtain the asymptotic rate approximation.

\subsection{Asymptotic Rate Approximation}\label{rate}
We now investigate the asymptotic approximation for the exact sum rate whose closed form expression has been derived in Section \ref{exact}. Two
additional issues arise in the heterogeneous multicell setting under partial feedback when compared with the standard homogeneous setting under
full feedback.

The first issue regards the heterogeneous statistics of the $\mathsf{SINR}$ for different users. In the homogeneous setting, the maximization or
the order statistics is over the same CDF. Recall that the use of CDF-based scheduling in this paper has enabled each user to virtually feel
that the other associated users are experiencing the same CDF for scheduling competition. Therefore, for a given user $k$, the order statistics
is over the CDF of  user $k$'s received CQI, which makes the individual user rate more interesting than the sum rate.

The second issue arises due to the effect of partial feedback. In the full feedback case, the number of CQI values to maximize over at the
scheduler is fixed and equals the number of the associated users $K$. However, due to partial feedback, the number of CQI values to maximize
over at the scheduler is a random quantity. In other words, partial feedback results in a random effect on multiuser diversity. In the exact
analysis in Section \ref{scheduling}, this effect is reflected in the use of $\mathcal{U}_{n,M}$. We are interested in the asymptotic effect
when the number of users grows large. To examine this random effect in the asymptotic analysis, denote the sequence of random variables
$\kappa_n(K)$ as the number of CQI values fed back for resource block $n$ with $K$ associated users. It is easy to see from (\ref{system_eq_3})
and (\ref{system_eq_6}) that $\kappa_n(K)$ are binomial distributed with probability of success $\frac{M}{N}$ under best-M partial feedback.
Thus by the strong law of large numbers, as $K$ grows, the number of CQI values fed back for each resource block becomes $\frac{KM}{N}$.
Moreover, the convergence property of the sequence $\kappa_n(K)$ can be shown by invoking the central limit theorem \cite{casella01}:
\begin{equation}\label{asymptotic_eq_1}
\mathop{\lim}\limits_{K\rightarrow\infty}\sqrt{K}\left(\frac{\kappa_n(K)}{K}-\frac{M}{N}\right)\mathop{\rightarrow}^{d}\mathcal{N}\left(0,\frac{M}{N}\left(1-\frac{M}{N}\right)\right),
\end{equation}
where $d$ indicates convergence in distribution. Therefore, by employing the techniques which study the extremes over random sample size
\cite{galambos78, berman62}, we have the following lemma.
\begin{lemma}\label{lemma_5}
When the number of associated users $K$ goes large, the extreme order statistics \cite{galambos78} of the received CQI for a given user $k$ can
be efficiently approximated by $\left(F_{Y_{k,M}^{(0)}}\right)^{\frac{KM}{N}}$.
\end{lemma}
\begin{proof}
The proof is given in Appendix \ref{appenB}.
\end{proof}

Now consider the limiting distribution of the maximum rate in order to derive the asymptotic approximation for the exact rate. Specially, we
examine the limiting distribution of the rate $R_{k,M}$,
\begin{equation}\label{asymptotic_eq_2}
R_{k,M}=T(Y_{k,M})=\log_2(1+Y_{k,M}),
\end{equation}
where the superscript $(0)$ is temporally dropped for representation simplicity, and will be added later to tailor the results for specific
$K_0$ and $Y_{k,M}^{(0)}$. Note that the function $T(\cdot)$ in (\ref{asymptotic_eq_2}) makes it tedious to directly check the conditions needed
to enable finding the form of the asymptotic distribution. In \cite{song06}, a limiting throughput distribution theorem is proposed for the full
feedback single cell case for a narrowband system. Herein, we generalize the result to be applicable to the general $\mathsf{SINR}$ case in the
general partial feedback OFDMA scenario with the following best-M limiting throughput distribution (LTD-M) theorem.

\begin{theorem} \label{theorem_3}
(\textit{LTD-M Theorem}) Assume that under the best-M partial feedback strategy with $N$ resource blocks and $K$ associated users, the CQI
received at the scheduler for user $k$, $Y_{k,M}$ is a nonnegative random variable with CDF $F_{Y_{k,M}}(x)$ such that
$f_{Y_{k,M}}(x)=F_{Y_{k,M}}'(x)>0$ and $\omega(F_{Y_{k,M}})\triangleq\sup\{x:F_{Y_{k,M}}(x)<1\}=\infty$. If
$\mathop{\lim}\limits_{x\rightarrow\infty}\frac{xf_{Y_{k,M}}(x)}{1-F_{Y_{k,M}}(x)}=\phi>0$, $F_{Y_{k,M}}\in\mathcal{D}(G_1)$, i.e.,
$F_{Y_{k,M}}$ belongs to the domain of attraction of the Fr{\'e}chet distribution, or if
$\mathop{\lim}\limits_{x\rightarrow\infty}\frac{d}{dx}\left[\frac{1-F_{Y_{k,M}}(x)}{f_{Y_{k,M}}(x)}\right]=0$, $F_{Y_{k,M}}\in\mathcal{D}(G_3)$,
i.e., $F_{Y_{k,M}}$ belongs to the domain of attraction of the Gumbel distribution, then the distribution of the throughput for user $k$,
$F_{R_{k,M}}(r)=F_{Y_{k,M}}(T^{-1}(r))\in\mathcal{D}(G_3)$, i.e., $F_{R_{k,M}}$ belongs to the domain of attraction of the Gumbel distribution.
Moreover, the normalizing constants \cite{galambos78} for user $k$ are given by
\begin{align}
a_{k:K}(M)&=\log_2\left(1+F_{Y_{k,M}}^{-1}\left(1-\frac{N}{KM}\right)\right),\notag\\
\label{asymptotic_eq_3} b_{k:K}(M)&=\log_2\left(\frac{1+F_{Y_{k,M}}^{-1}\left(1-\frac{N}{KMe}\right)}{1+
F_{Y_{k,M}}^{-1}\left(1-\frac{N}{KM}\right)}\right).
\end{align}
\end{theorem}
\begin{proof}
The proof is given in Appendix \ref{appenB}.
\end{proof}
\emph{Remark:} The LTD-M theorem enables us to study the distribution of $Y_{k,M}$ instead of directly examining $F_{R_{k,M}}$. Also, note that
the relationship of the type of convergence between $Y_{k,M}$ and $Z_k$ has been revealed in Theorem \ref{theorem_2}. Thus the connection
between $Z_k$ and $R_{k,M}$ can be established by combining the two theorems.

The normalizing constants in Theorem \ref{theorem_3} can be used to obtain the asymptotic rate approximation. Denote $\mathcal{C}_k^{(0)}(M)$ as
the asymptotic approximation for the individual rate of user $k$ in cell $B_0$ with total associated users $K_0.$ Then according to the property
that convergence in distribution for the maximum nonnegative random variables results in moment convergence \cite{pickands68, song06},
$\mathcal{C}_k^{(0)}(M)$ can be evaluated by the normalizing constants as follows\footnote[9]{This form of asymptotic approximation leverages
the first and second order moments of the extreme order statistics. Dealing with higher order moments and eventually the rate of convergence is
left to our future work.}
\begin{equation}
\label{asymptotic_eq_4}\mathcal{C}_k^{(0)}(M)=\frac{1}{K_0}\left(1-\left(1-\frac{M}{N}\right)^{K_0}\right)
\left(a_{k:K_0}^{(0)}(M)+E_0b_{k:K_0}^{(0)}(M)\right),
\end{equation}
where $E_0$ is the Euler constant, and $(1-\frac{M}{N})^K$ is the probability of scheduling outage. According to the CDF-based scheduling
policy, the asymptotic approximation for the sum rate, denoted by $\mathcal{C}^{(0)}(M)$, can be computed as
\begin{equation}
\label{asymptotic_eq_5}\mathcal{C}^{(0)}(M)=\frac{1}{K_0}\left(1-\left(1-\frac{M}{N}\right)^{K_0}\right)\sum_{k=1}^{K_0}\left(a_{k:K_0}^{(0)}(M)+E_0b_{k:K_0}^{(0)}(M)\right).
\end{equation}
The form in (\ref{asymptotic_eq_5}) is simpler than the exact analytic expression derived in Section \ref{exact} and can be an alternate basis
for studying heterogeneous networks. Looking again at the normalizing constants in (\ref{asymptotic_eq_3}), the specific expressions  involve
the inverse of the distribution function, $F_{Y_{k,M}}^{-1}(\cdot)$. In general, due to the complicated form of the $\mathsf{SINR}$ as well as
the procedure to evaluate $F_{Y_{k,M}}$, this inverse function can not be expressed in simple closed form except in some simplified cases. Since
the CDF is a function of a scalar and is monotonically increasing, standard iterative algorithms are well suited for its computation. Now we
consider the two aforementioned simplified cases: the one-dominant interference limited case and the noise limited case for illustration. For
these two special cases with full feedback and best-1 feedback, the inverse of the distribution function can be computed in closed form, which
are summarized in the following corollaries.
\begin{corollary} \label{corollary_3}
In the one-dominant interference limited case under full feedback, the specific form for the normalizing constants are given by:
\begin{align}
a_{k:K_0}^{(0)}(N)&=\log_2\left(1+\frac{\rho_k^{(0)}}{\rho_k^{(1)}}(K_0-1)\right),\notag\\
\label{asymptotic_eq_6}b_{k:K_0}^{(0)}(N)&=\log_2\left(\frac{1+\frac{\rho_k^{(0)}}{\rho_k^{(1)}}(K_0e-1)}{1+\frac{\rho_k^{(0)}}{\rho_k^{(1)}}(K_0-1)}\right).
\end{align}
In the noise limited case under full feedback, the specific form for the normalizing constants are given by:
\begin{align}
a_{k:K_0}^{(0)}(N)&=\log_2\left(1+\rho_k^{(0)}\ln K_0\right),\notag\\
\label{asymptotic_eq_7}b_{k:K_0}^{(0)}(N)&=\log_2\left(1+\frac{\rho_k^{(0)}}{1+\rho_k^{(0)}\ln K_0}\right).
\end{align}
\end{corollary}
\begin{proof}
The proof is given in Appendix \ref{appenB}.
\end{proof}

\begin{corollary} \label{corollary_4}
In the one-dominant interference limited case under best-1 feedback, the specific form for the normalizing constants are given by:
\begin{align}
a_{k:K_0}^{(0)}(1)&=\log_2\left(1+\frac{\rho_k^{(0)}}{\rho_k^{(1)}}\frac{(K_0-N)^{\frac{1}{N}}}{K_0^{\frac{1}{N}}-(K_0-N)^{\frac{1}{N}}}\right),\notag\\
\label{asymptotic_eq_8}b_{k:K_0}^{(0)}(1)&=\log_2\left(\frac{1+\frac{\rho_k^{(0)}}{\rho_k^{(1)}}\frac{(K_0e-N)^{\frac{1}{N}}}{(K_0e)^{\frac{1}{N}}-(K_0e-N)^{\frac{1}{N}}}}{1+\frac{\rho_k^{(0)}}{\rho_k^{(1)}}\frac{(K_0-N)^{\frac{1}{N}}}{K_0^{\frac{1}{N}}-(K_0-N)^{\frac{1}{N}}}}\right).
\end{align}
In the noise limited case under best-1 feedback, the specific form for the normalizing constants are given by:
\begin{align}
a_{k:K_0}^{(0)}(1)&=\log_2\left(1+\rho_k^{(0)}\ln\frac{K_0^{\frac{1}{N}}}{K_0^{\frac{1}{N}}-(K_0-N)^{\frac{1}{N}}}\right),\notag\\
\label{asymptotic_eq_9}b_{k:K_0}^{(0)}(1)&=\log_2\left(\frac{1+\rho_k^{(0)}\ln\frac{(K_0e)^{\frac{1}{N}}}{(K_0e)^{\frac{1}{N}}-(K_0e-N)^{\frac{1}{N}}}}{1+\rho_k^{(0)}\ln\frac{K_0^{\frac{1}{N}}}{K_0^{\frac{1}{N}}-(K_0-N)^{\frac{1}{N}}}}\right).
\end{align}
\end{corollary}
\begin{proof}
The proof is given in Appendix \ref{appenB}.
\end{proof}

\begin{figure}[t]
\centering
\begin{tabular}{cc}
    \includegraphics[width=0.5\linewidth]{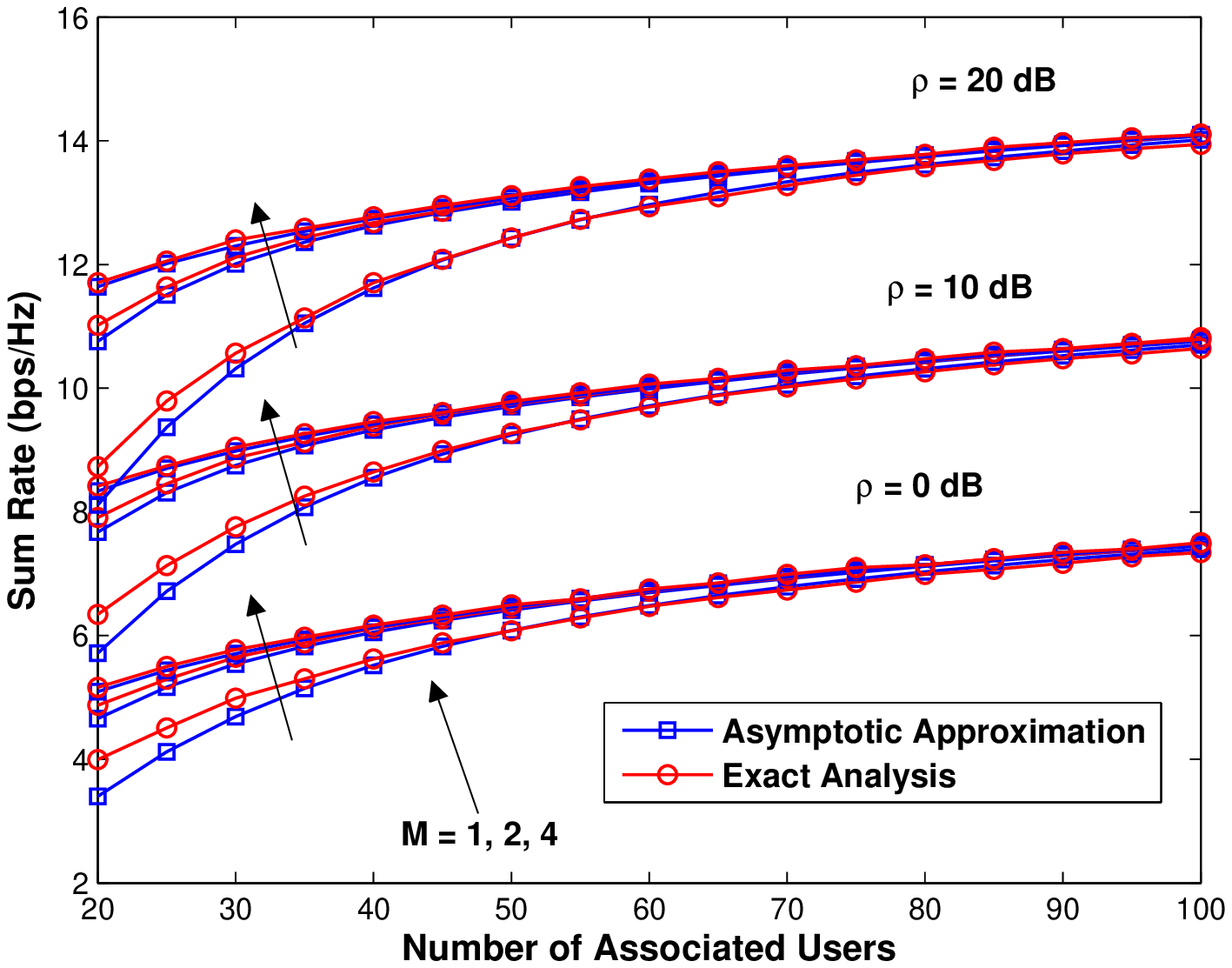}&
    \includegraphics[width=0.5\linewidth]{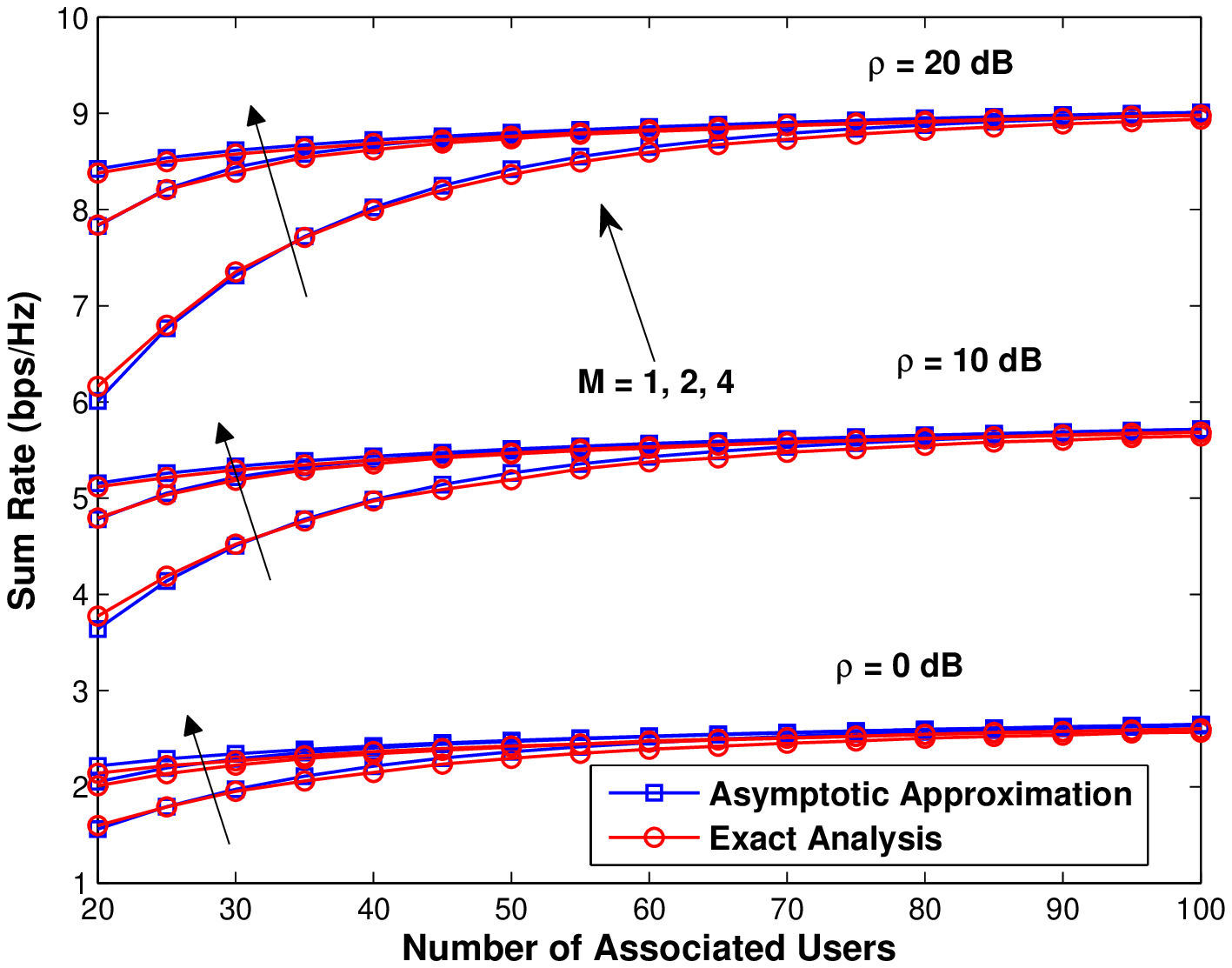}\\
    \scriptsize{(a)}&
    \scriptsize{(b)}\\
\end{tabular}
\caption{Comparison of the sum rate for best-M feedback obtained using the exact analysis and the asymptotic analysis under different symmetric
large scale effects for different $M$ with respect to the number of users ($N=16$; $M=1,2,4$): (a) the one-dominant interference limited case
($\rho\triangleq \frac{\rho^{(0)}}{\rho^{(1)}}$); (b) the noise limited case ($\rho\triangleq\rho^{(0)}$).} \label{fig_2}
\end{figure}

\begin{table}[t]\footnotesize
\caption{The Main Steps of the Analytical Framework for the Exact Analysis and the Asymptotic Analysis} \label{tab_1} \centering

\begin{tabular}{|c||c|c||c|c|}
\hline  \multirow{2}{*}{Framework} & \multicolumn{2}{|c||}{Exact Analysis} & \multicolumn{2}{|c|}{Asymptotic Analysis}\\
\cline{2-5} & Technical Steps & Main Results & Technical Steps & Main Results\\
\hline Step $1$ & Statistics of $Z_k^{(0)}$ & & Type of Convergence of $F_{Z_k^{(0)}}$ & Corollary $1$\\
\cline{1-2} \cline{4-5}
Step $2$ & Statistics of $Y_{k,M}^{(0)}$ and $\tilde{Y}_{k,M}^{(0)}$ & Lemma $1-3$ & Type of Convergence of $F_{Y_{k,M}^{(0)}}$ & Theorem $2$\\
\cline{1-2} \cline{4-5}
Step $3$ & Statistics of $X_{M}^{(0)}\mid k^*=k,|\mathcal{U}_{M}|=\tau_0$ & Theorem $1$ & Type of Convergence of $F_{R_{k,M}^{(0)}}$ & Theorem $3$\\
\cline{1-2} \cline{4-5}
Step $4$ & $C^{(0)}(M)$ and $C_k^{(0)}(M)$ & & $\mathcal{C}^{(0)}(M)$ and $\mathcal{C}_k^{(0)}(M)$ & Theorem $3$\\
\hline
\end{tabular}

\end{table}

For the general best-M partial feedback case, the normalizing constants can be obtained using (\ref{asymptotic_eq_3}) and the specific CDF
for the corresponding case. To illustrate the benefit of asymptotic analysis for the two simplified cases, we conduct a numerical study to
compare the sum rate obtained using the exact analysis and the asymptotic one under different symmetric large scale effects in Fig. \ref{fig_2}.
It is interesting to note that the asymptotic expressions hold well even for small number of users, which means the convergence to the limiting
distribution is fast.

Up to now, we have leveraged exact analysis and asymptotic analysis to derive useful closed form results for the exact sum rate and the
asymptotic approximation. For easy reference, the main technical steps and the main results are summarized in Table \ref{tab_1}. In the next
part, the procedure to determine the minimum required partial feedback is examined.

\subsection{Determining the Minimum Required Partial Feedback}\label{determine}
Firstly, the asymptotic optimality of the best-1 feedback is presented in the following corollary.
\begin{corollary} \label{corollary_5}
When the number of associated users $K\rightarrow\infty$, the performance loss of using best-1 feedback in terms of sum rate approaches zero.
\end{corollary}
\begin{proof}
The proof is given in Appendix \ref{appenB}.
\end{proof}
We are more interested in the pre-asymptotic user regime where the simple best-1 strategy is no longer optimal. The goal is to choose the minimum
required M without seriously degrading system sum rate when compared to a system with full feedback. The selection of M can
be formulated as the solution to the following optimization problem:
\begin{equation}
\label{asymptotic_eq_10} \mathrm{Find\; the\; minimum\;} M^*,\quad s.t.\; \frac{C^{(0)}(M^*)}{C^{(0)}(N)}\geq \eta,
\end{equation}
where $C^{(0)}(M)$ refers to the exact analytic expression derived in Section \ref{exact} and $\eta$ is a system defined threshold. As
mentioned previously, the established asymptotic expressions are more
computationally efficient than the exact one. By leveraging the asymptotic
approximation, the problem (\ref{asymptotic_eq_10}) can be reformulated as
\begin{equation}
\label{asymptotic_eq_11} \mathrm{Find\; the\; minimum\;} \tilde{M}^*,\quad s.t.\;
\frac{\mathcal{C}^{(0)}(\tilde{M}^*)}{\mathcal{C}^{(0)}(N)}\geq \eta,
\end{equation}
where $\mathcal{C}^{(0)}(M)$ refers to the asymptotic approximation in (\ref{asymptotic_eq_5}).

We see from (\ref{asymptotic_eq_10}) and (\ref{asymptotic_eq_11}) that $M^*$ or $\tilde{M}^*$ depends on the number of users associated with the
base station and the corresponding large scale channel effects. Since these factors can be highly diverse in a heterogeneous network, the
minimum required partial feedback is inherently different across different cells. Therefore, by employing the established asymptotic results,
$\tilde{M}^*$ can be quickly determined to track $M^*$ in order to design situational-aware heterogeneous partial feedback.

\section{Numerical Results}\label{numerical}
In this section, we conduct a numerical study to support our analysis. A simple heterogeneous network is modeled with two macrocells each with
two picocells inside. The locations of the picocells are randomly placed and then fixed for simulation. The system bandwidth is $5$ MHz, the
noise power spectral density is $-170$ dBm/Hz, and the number of resource blocks $N=16$. The transmit powers of the macrocell and picocell are
$43$ dBm and $30$ dBm respectively. The path loss (in dB) model in \cite{3gpp10} with $2$ GHz central frequency is employed: the path loss from
the macrocell base station to users is $15.3+37.6\log_{10} d$ for distance $d$ in meters; the path loss from the picocell base station to users
is $30.6+36.7\log_{10} d$ for distance $d$ in meters. Log-normal shadowing is assumed with standard deviation of $8$ dB. The radius of the
macrocell and picocell is assumed to be $500$ m and $100$ m respectively. For each drop in simulation, users are randomly placed and the cell
association is determined by the large scale effects and fixed.

\begin{figure}[t]
\centering
\begin{tabular}{cc}
    \includegraphics[width=0.5\linewidth]{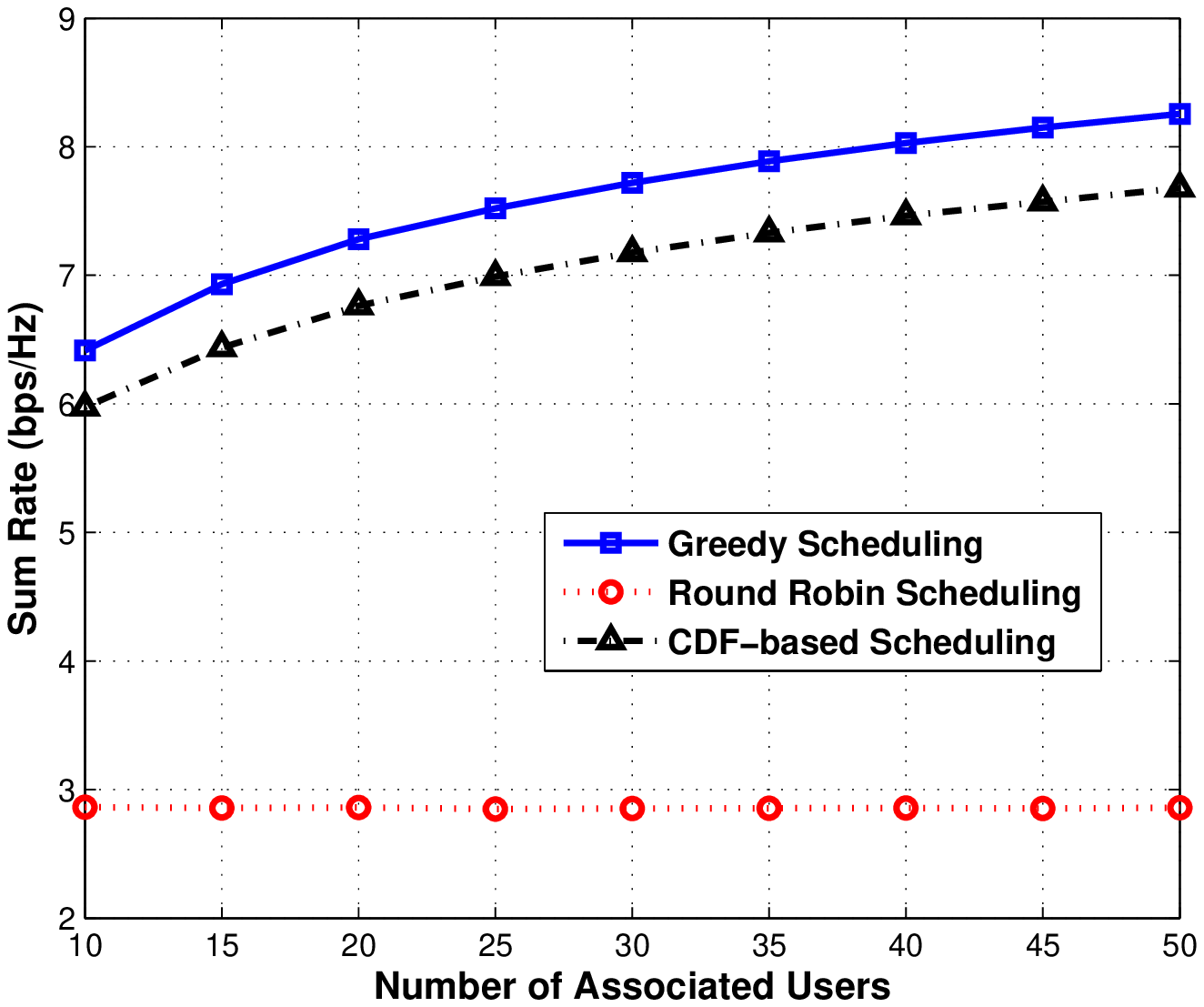}&
    \includegraphics[width=0.5\linewidth]{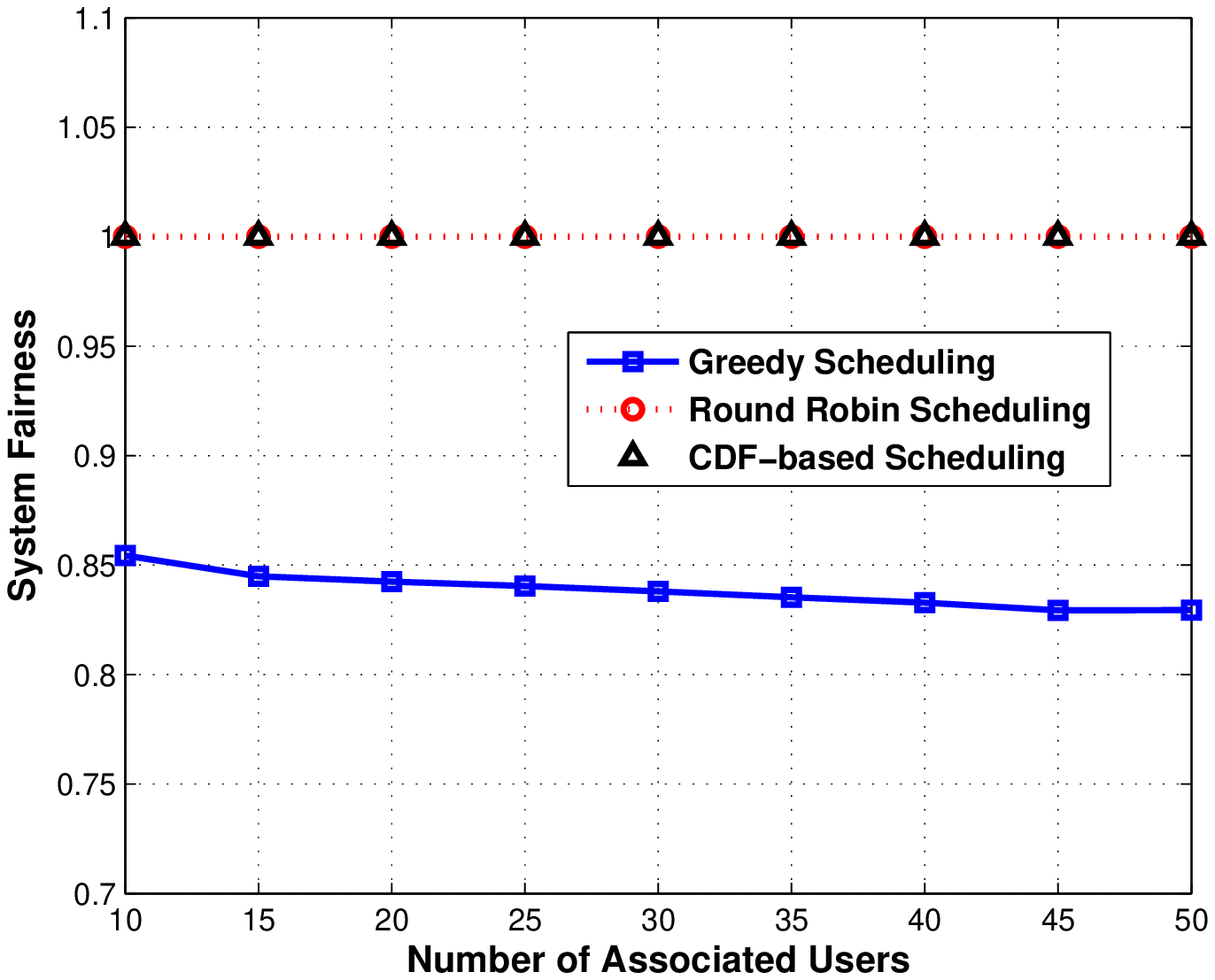}\\
    \scriptsize{(a)}&
    \scriptsize{(b)}\\
\end{tabular}
\caption{Comparison of the three scheduling policies: the greedy policy, the round robin policy, and the CDF-based policy under best-M partial
feedback strategy with respect to the associated users in a macrocell ($N=16$; $M=4$): (a) The sum rate comparison; (b) The system fairness
comparison.} \label{fig_3}
\end{figure}

Firstly, the performance of CDF-based scheduling is compared with the greedy scheduling policy and the round robin scheduling policy in Fig.
\ref{fig_3}. In this simulation, users are assumed to employ the best-M partial feedback with $M=4$. In Fig. \ref{fig_3} (a), the sum rate with
respect to the number of associated users in a macrocell is shown, which is averaged by performing $1000$ independent drops. It can be seen that
the round robin policy does not invoke multiuser diversity at all, and the sum rate performance of the CDF-based policy is close to the greedy
policy. Fig. \ref{fig_3} (b) compares the system fairness for the three policies. The system fairness $\Theta$ is defined and discussed in
\cite{elliott02, yang06} using the following form: $\Theta\triangleq -\sum_{k=1}^K \mathbb{P}_k\frac{\ln(\mathbb{P}_k)}{\ln K}$, where
$\mathbb{P}_k$ refers to the proportion of resources assigned to user $k$ with the normalization factor $\ln K$. The system fairness for the
round robin policy and the CDF-based policy is $1$ despite the number of associated users and the heterogeneous channel effects. However, the
system fairness for the greedy policy is decreasing when more users are associated. This is due to the fact that some high geometry users occupy
the system resources with high probability when the greedy system has more serving users. Therefore, the CDF-based scheduling policy enjoys the
multiuser diversity while guaranteeing fairness at the same time, which makes it well suited for the heterogeneous framework.

\begin{figure}[t]
\centering
    \includegraphics[width=0.7\linewidth]{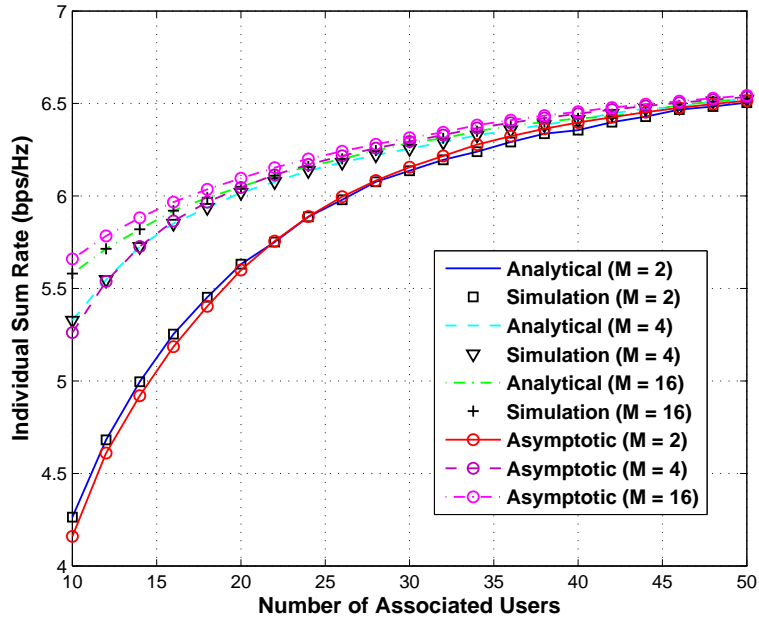}
\caption{Comparison of the individual sum rate (individual user rate multiplied by the number of associated users) obtained from simulation,
analytical result, and asymptotic approximation for different partial feedback $M$ with respect to the number of associated users for a randomly
selected user in macrocell ($N=16$; $M=2,4,16$).} \label{fig_4}
\end{figure}

Next, in order to evaluate the derived closed form results from exact analysis and the corresponding asymptotic approximation, one individual
user is randomly selected for demonstration. To illustrate the scaling performance with respect to the number of associated users, the so called
individual sum rate for this user is of interest. The individual sum rate is the individual user rate multiplied by the number of associated
users. Fig. \ref{fig_4} plots the individual sum rate obtained from the analytical expression by exact analysis, from simulation, and from the
established asymptotic approximation using normalizing constants. Different partial feedback cases are also shown for comparison. It can be
observed that the analytical expression and the simulation results are in full agreement. Also, the asymptotic approximation tracks the system
performance very well. Furthermore, the rate gap between the partial feedback case and full feedback case becomes negligible when the number of
associated users increases.

\begin{figure}[t]
\centering
\begin{tabular}{cc}
    \includegraphics[width=0.5\linewidth]{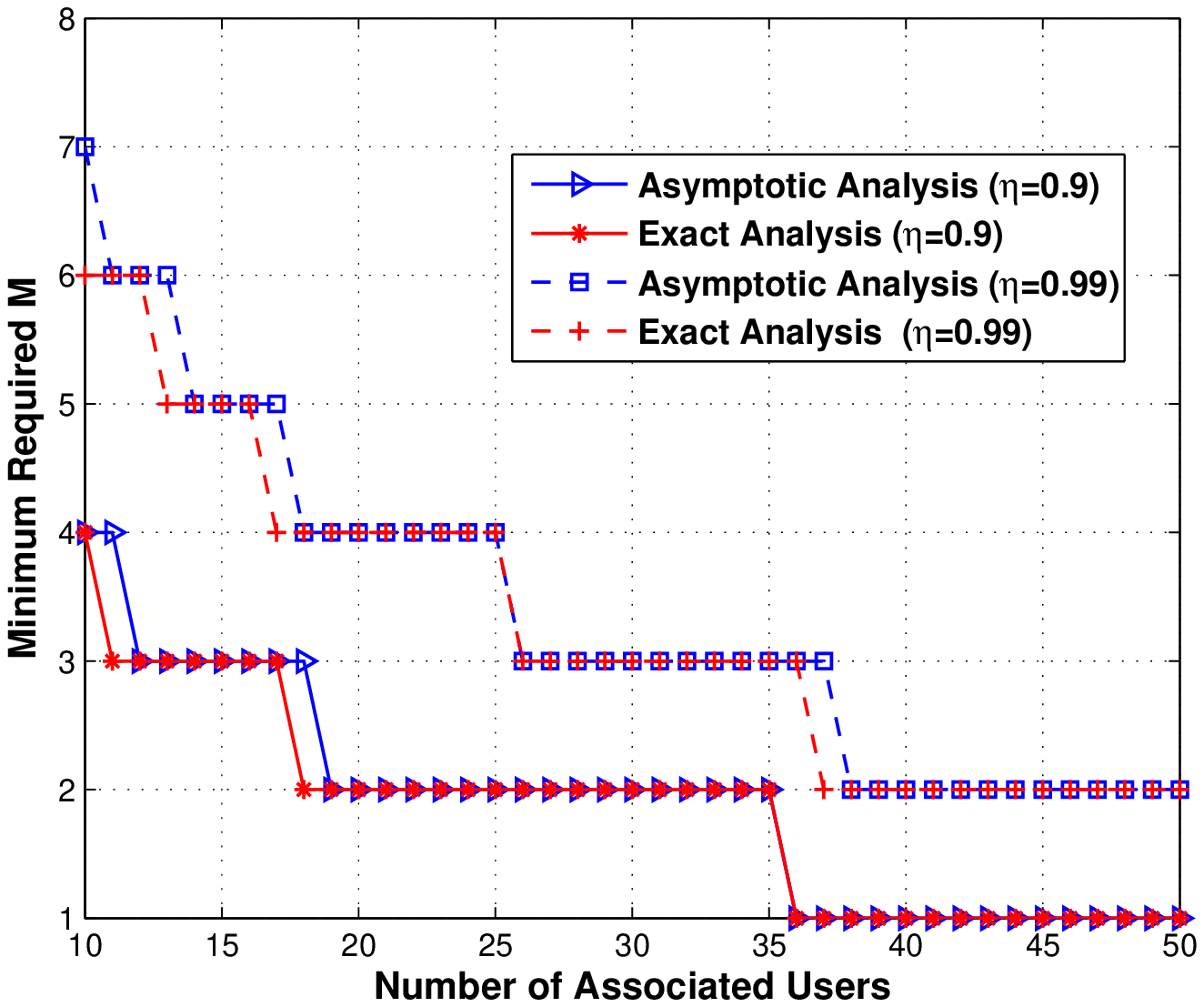}&
    \includegraphics[width=0.5\linewidth]{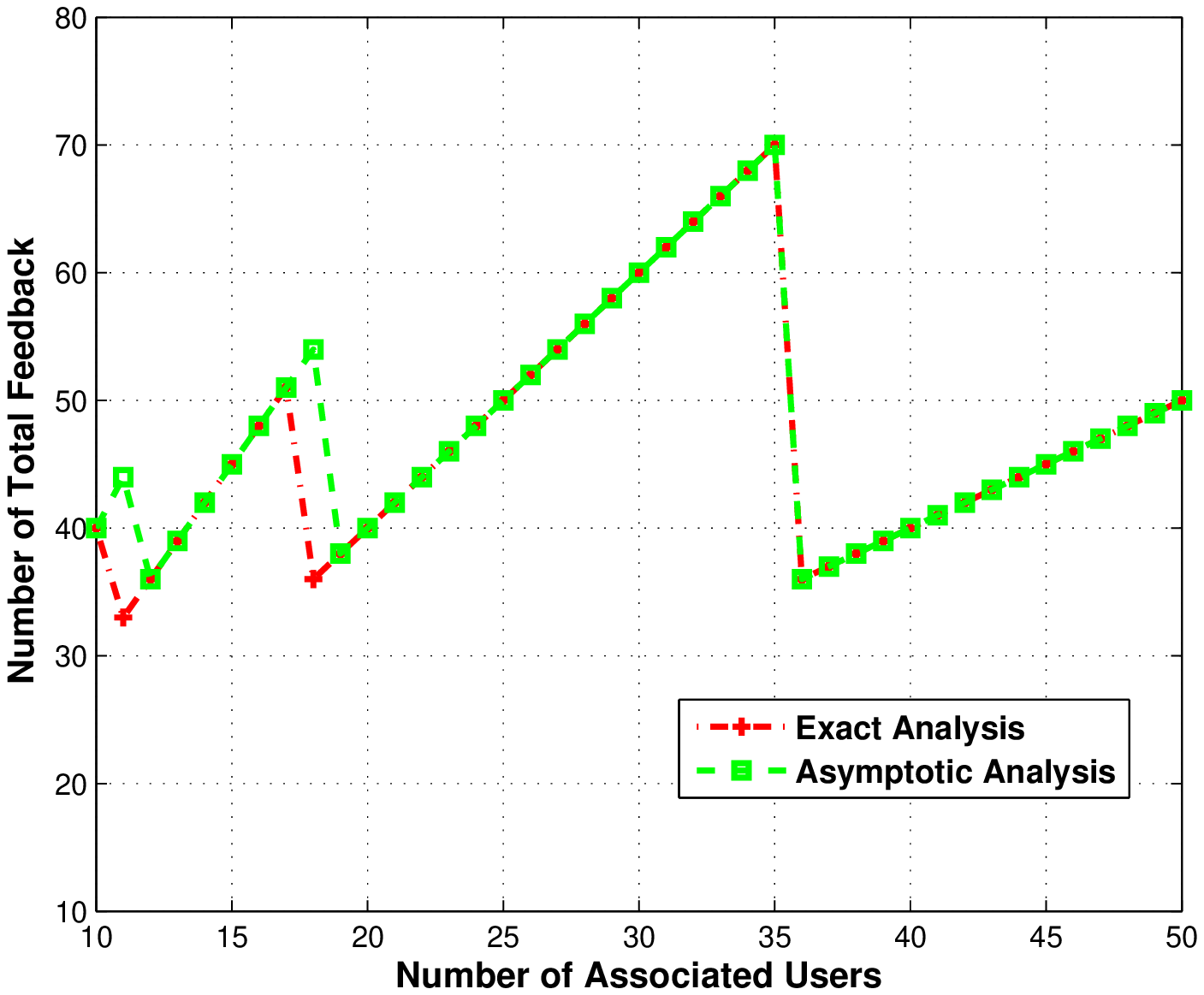}\\
    \scriptsize{(a)}&
    \scriptsize{(b)}\\
\end{tabular}
\caption{Partial feedback with respect to the number of associated users for a macrocell ($N=16$): (a) Comparison of the minimum required $M$
obtained from exact analysis and asymptotic analysis under different thresholds ($\eta=0.9,0.99$); (b) Comparison of the total number of
required partial feedback from exact analysis and asymptotic analysis under threshold $\eta=0.9$.} \label{fig_5}
\end{figure}

Finally, Fig. \ref{fig_5} (a) examines the minimum required partial feedback $M^*$ obtained by using the expression for the sum rate from the
exact analysis and the asymptotic approximation for a macrocell. Two different thresholds are set for evaluation: $\eta=0.9, 0.99$. It can be
seen that the results obtained using asymptotic analysis track the results from exact analysis very well, especially for lower threshold and
larger number of users. Since the number of users associated with each cell as well as the large scale channel effects can lie in diverse
ranges, this results in heterogeneous partial feedback in heterogeneous multicell networks. The total number of partial feedback with respect to
the number of associated users for a macrocell is illustrated in Fig. \ref{fig_5} (b) under the threshold $\eta=0.9$. The total number of
partial feedback is calculated by multiplying the minimum required partial feedback $M^*$ with a given number of associated users. It can be
seen that the range of variation for the total number of partial feedback is limited. Even though the number of associated users is $5$ times
larger, the total number of partial feedback does not change much. This is due to the heterogeneous partial feedback design to adapt the number
of partial feedback to the number of users as well as the channel conditions.

\section{Conclusion}\label{conclusion}
In this paper, an analytical framework is proposed and developed to investigate the performance of situational-aware heterogeneous partial
feedback in an OFDMA-based heterogeneous multicell using the best-M partial feedback strategy. The system model is general and thus the obtained
results can be generalized and applied to conduct system evaluation with alternate statistical models. The CDF-based scheduling policy employed
in this paper has the desired property of supporting multiuser diversity while maintaining scheduling fairness among the contending users to
guarantee each user's data rate despite of different locations and large scale channel effects. The exact closed form sum rate is obtained for
the multicell model by suitable decomposition and expansion of the received CQI at the scheduler side. Asymptotic analysis is carried out to
draw further insight into the multicell model with partial feedback. Interestingly, the effect of partial feedback does not alter the type of
convergence of the received CQI. The random effect of multiuser diversity caused by partial feedback is also examined and asymptotic
approximations are derived by utilizing the normalizing constants. The established asymptotic approximation tracks the exact system performance
well even for small number of users. Therefore, it can be leveraged to quickly determine the minimum required partial feedback in a given cell.
In this paper, only a single antenna is assumed and the scheduling policy is fully distributed. Our future work will consider a multi-antenna
setting and more advanced scheduling policies.


%

\appendices
\section{}\label{appenA}
\textit{Proof of Lemma \ref{lemma_1}:} Denote $\zeta=\sum_{b=1}^{J_k}\rho_k^{(b)}|H_{k,n}^{(b)}|^2$, which is a weighted sum of $\chi^2(2)$
random variable. In practical system setting, the large scale effects from the interfering cells are distinct, the PDF of $\zeta$ is derived to
be \cite{tokgoz06, bjornson09}:
\begin{equation}
\label{appen:eq_1} f_{\zeta}(\zeta)=\left(\sum_{b=1}^{J_k}\frac{\varpi_k^{(b)}}{\rho_k^{(b)}}e^{-\frac{\zeta}{\rho_k^{(b)}}}\right)u(x),
\end{equation}
where $\varpi_k^{(b)}=\mathop{\prod}\limits_{\substack{i=1\\i\neq b}}\frac{\rho_k^{(b)}}{\rho_k^{(b)}-\rho_k^{(i)}}$. Then the PDF of
$Z_k^{(0)}$ can be obtained as follows
\begin{align}
f_{Z_k^{(0)}}(x)&=\int_0^{\infty}f_{Z_k^{(0)}|\zeta}(x|\zeta)f_{\zeta}(\zeta)d\zeta\notag\\
&=\frac{1}{\rho_k^{(0)}}e^{-\frac{x}{\rho_k^{(0)}}}\mathop{\sum}\limits_{b=1}^{J_k}\frac{\varpi_k^{(b)}}{\rho_k^{(b)}}\int_0^{\infty}(1+\zeta)e^{-\left(\frac{x}{\rho_k^{(0)}}+\frac{1}{\rho_k^{(b)}}\right)\zeta}d\zeta\notag\\
\label{appen:eq_2}
&=\frac{1}{\rho_k^{(0)}}e^{-\frac{x}{\rho_k^{(0)}}}\mathop{\sum}\limits_{b=1}^{J_k}\frac{\varpi_k^{(b)}}{\rho_k^{(b)}}\left(\frac{\rho_k^{(0)}}{x+\frac{\rho_k^{(0)}}{\rho_k^{(b)}}}+\frac{\left(\rho_k^{(0)}\right)^2}{\left(x+\frac{\rho_k^{(0)}}{\rho_k^{(b)}}\right)^2}\right)u(x).
\end{align}

\medskip

\textit{Proof of Lemma \ref{lemma_2}:} It is clear that
$d(F_{Z_k^{(0)}}(x))^{\epsilon}=\epsilon(F_{Z_k^{(0)}}(x))^{\epsilon-1}f_{Z_k^{(0)}}(x)dx$. Employing the binomial theorem \cite{stark02}, and
substituting the expressions for $F_{Z_k^{(0)}}(x)$ and $f_{Z_k^{(0)}}(x)$ yield
\begin{align}
d(F_{Z_k^{(0)}}(x))^{\epsilon} &=\epsilon\sum_{\ell=0}^{\epsilon-1}{\epsilon-1\choose
\ell}(-1)^{\ell}\left(\sum_{b=1}^{J_k}\frac{\varpi_k^{(b)}e^{-\frac{x}{\rho_k^{(0)}}}\rho_k^{(0)}}{\rho_k^{(0)}+\rho_k^{(b)}x}\right)^{\ell}\left(\sum_{b=1}^{J_k}\frac{\varpi_k^{(b)}e^{-\frac{x}{\rho_k^{(0)}}}}{\rho_k^{(0)}+\rho_k^{(b)}x}+\frac{\rho_k^{(0)}\rho_k^{(b)}\varpi_k^{(b)}e^{-\frac{x}{\rho_k^{(0)}}}}{\left(\rho_k^{(0)}+\rho_k^{(b)}x\right)^2}\right)dx\notag\\
\label{appen:eq_3} & \mathop{=}\limits^{(a)} \epsilon\sum_{\ell=0}^{\epsilon-1}{\epsilon-1\choose
\ell}\frac{(-1)^{\ell}}{\ell+1}d\left(1-\left(\sum_{b=1}^{J_k}\frac{\varpi_k^{(b)}e^{-\frac{x}{\rho_k^{(0)}}}\rho_k^{(0)}}{\rho_k^{(0)}+\rho_k^{(b)}x}\right)^{\ell+1}\right),
\end{align}
where (a) follows from the differentiation property of $(1-F_{Z_k^{(0)}}(x))^{\ell+1}$.

\medskip

\textit{Proof of Lemma \ref{lemma_3}:} Applying the multinomial theorem \cite{stark02} yields
\begin{equation}\label{appen:eq_4}
\left(\sum_{b=1}^{J_k}\varpi_k^{(b)}\frac{\rho_k^{(0)}}{\rho_k^{(b)}}\frac{1}{x+\frac{\rho_k^{(0)}}{\rho_k^{(b)}}}\right)^{\ell+1}=\sum_{j_1+\cdots+j_{J_k}=\ell+1}{\ell+1
\choose
j_1,\ldots,j_{J_k}}\mathop{\prod}\limits_{b=1}^{J_k}\frac{\left(\frac{\varpi_k^{(b)}\rho_k^{(0)}}{\rho_k^{(b)}}\right)^{j_b}}{{\left(x+\frac{\rho_k^{(0)}}{\rho_k^{(b)}}\right)^{j_b}}}.
\end{equation}
Exploiting the partial fraction expansion \cite{gradshteyn07} generates the expanded form in (\ref{exact_eq_4}) with specific expression for
$\psi_{k,i}^{(b)}$ defined in (\ref{exact_eq_5}).

For illustration purpose, the expansion for $J_k=2$ case, which corresponds to two major interferers is now presented. In this case, applying
binomial theorem is sufficient for expansion, which yields
\begin{align}
&\left(\frac{\varpi_k^{(1)}\rho_k^{(0)}}{\rho_k^{(1)}}\frac{1}{x+\frac{\rho_k^{(0)}}{\rho_k^{(1)}}}+\frac{\varpi_k^{(2)}\rho_k^{(0)}}{\rho_k^{(2)}}\frac{1}{x+\frac{\rho_k^{(0)}}{\rho_k^{(2)}}}\right)^{\ell+1}\notag\\
&=\sum_{j=0}^{\ell+1}{\ell+1\choose
j}\left(\rho_k^{(0)}\right)^{\ell+1}\left(\frac{\varpi_k^{(1)}}{\rho_k^{(1)}}\right)^{\ell+1-j}\left(\frac{\varpi_k^{(2)}}{\rho_k^{(2)}}\right)^j\frac{1}{\left(x+\frac{\rho_k^{(0)}}{\rho_k^{(1)}}\right)^{\ell+1-j}\left(x+\frac{\rho_k^{(0)}}{\rho_k^{(2)}}\right)^j}\notag\\
\label{appen:eq_23}&=\sum_{j=0}^{\ell+1}{\ell+1\choose
j}\left(\rho_k^{(0)}\right)^{\ell+1}\left(\frac{\varpi_k^{(1)}}{\rho_k^{(1)}}\right)^{\ell+1-j}\left(\frac{\varpi_k^{(2)}}{\rho_k^{(2)}}\right)^j\left(\sum_{i=0}^{\ell+1-j}\frac{\psi_{k,i}^{(1)}}{\left(x+\frac{\rho_k^{(0)}}{\rho_k^{(1)}}\right)^i}+\sum_{i=0}^{j}\frac{\psi_{k,i}^{(2)}}{\left(x+\frac{\rho_k^{(0)}}{\rho_k^{(2)}}\right)^i}\right),
\end{align}
where $\psi_{k,i}^{(1)}=(-1)^{\ell+1-j-i}{\ell-i\choose
j-1}\frac{1}{\left(\frac{\rho_k^{(0)}}{\rho_k^{(2)}}-\frac{\rho_k^{(0)}}{\rho_k^{(1)}}\right)^{\ell+1-i}}$ and
$\psi_{k,i}^{(2)}=(-1)^{j-i}{\ell-i\choose
j-i}\frac{1}{\left(\frac{\rho_k^{(0)}}{\rho_k^{(1)}}-\frac{\rho_k^{(0)}}{\rho_k^{(2)}}\right)^{\ell+1-i}}$ for the nontrivial $i>0$ case.

\medskip

\textit{Proof of Theorem \ref{theorem_1}:} The outcomes of Step $1$ and Step $2$ lead to direct integration to calculate
$\mathcal{G}_k(\epsilon)$ as follows
\begin{align}
\mathcal{G}_k(\epsilon)&=\frac{\epsilon}{\ln2}\sum_{\ell=0}^{\epsilon-1}{\epsilon-1\choose
\ell}\frac{(-1)^{\ell}}{\ell+1}\int_0^{\infty}\ln(1+x)d\left(1-\left(\sum_{b=1}^{J_k}\frac{\varpi_k^{(b)}e^{-\frac{x}{\rho_k^{(0)}}}\rho_k^{(0)}}{\rho_k^{(0)}+\rho_k^{(b)}x}\right)^{\ell+1}\right)\notag\\
&\mathop{=}\limits^{(a)} \frac{\epsilon}{\ln2}\sum_{\ell=0}^{\epsilon-1}{\epsilon-1\choose
\ell}\frac{(-1)^{\ell}}{\ell+1}\sum_{j_1+\cdots+j_{J_k}=\ell+1}{\ell+1 \choose
j_1,\ldots,j_{J_k}}\sum_{b=1}^{J_k}\sum_{i=0}^{j_b}\psi_{k,i}^{(b)}\notag\\
\label{appen:eq_5}&\quad\times\mathop{\prod}\limits_{b=1}^{J_k}\left(\frac{\varpi_k^{(b)}\rho_k^{(0)}}{\rho_k^{(b)}}\right)^{j_b}\int_0^{\infty}\frac{e^{-\frac{(\ell+1)x}{\rho_k^{(0)}}}}{(1+x)(x+\frac{\rho_k^{(0)}}{\rho_k^{(b)}})^{i}}dx,
\end{align}
where (a) follows from integration by parts. The form of (\ref{exact_eq_6}) is expressed by the definition
$\mathcal{I}_1(\alpha,\beta,\gamma)\triangleq \int_0^{\infty}\frac{e^{-\alpha x}}{(1+x)(\beta+x)^{\gamma}} dx$. In order to compute
$\mathcal{I}_1(\alpha,\beta,\gamma)$ into closed form, firstly apply partial fraction expansion \cite{gradshteyn07} to
$\frac{1}{(1+x)(\beta+x)^{\gamma}}$, and then define the auxiliary integration $\mathcal{I}_2(\alpha,\beta,\gamma)\triangleq
\int_0^{\infty}\frac{e^{-\alpha x}}{(\beta+x)^{\gamma}} dx$. It is clear that when $\gamma=0$,
$\mathcal{I}_1(\alpha,\beta,\gamma)=\mathcal{I}_2(\alpha,1,1)$. For the non-trivial case when $\gamma\geq1$, employing the partial fraction
expansion and after some manipulation the following relationship is revealed between $\mathcal{I}_1(\cdot,\cdot,\cdot)$ and
$\mathcal{I}_2(\cdot,\cdot,\cdot)$:
\begin{equation}\label{appen:eq_6}
\mathcal{I}_1(\alpha,\beta,\gamma)=\frac{1}{(\beta-1)^{\gamma}}\mathcal{I}_2(\alpha,1,1)+\sum_{\imath=1}^{\gamma}\frac{(-1)^{\imath-1}}{(1-\beta)^\imath}\mathcal{I}_2(\alpha,\beta,\gamma-\imath+1).
\end{equation}
$\mathcal{I}_2(\alpha,\beta,\gamma)$ can be further computed by noting from \cite[3.352.2]{gradshteyn07} that
$\mathcal{I}_2(\alpha,\beta,1)=e^{\alpha\beta}E_1(\alpha\beta)$, where $E_1(x)=\int_x^\infty \frac{e^{-t}}{t}dt$ is the exponential integral
function of the first order \cite{abramowitz72}, and utilizing integration by parts. The closed form result for
$\mathcal{I}_2(\alpha,\beta,\gamma)$ is presented as follows
\begin{equation}
\label{appen:eq_7} \mathcal{I}_2(\alpha,\beta,\gamma)=\left\{
\begin{array}{l} \frac{(-1)^{\gamma-1}\alpha^{\gamma-1}e^{\alpha\beta}E_1(\alpha\beta)}{(\gamma-1)!}+\mathop{\sum}\limits_{\imath=1}^{\gamma-1}\frac{(\imath-1)!}{(\gamma-1)!}(-1)^{\gamma-\imath-1}\alpha^{\gamma-\imath-1}, \quad \gamma\geq2\\
e^{\alpha\beta}E_1(\alpha\beta), \quad \gamma=1\\\end{array} \right.\\
\end{equation}

\section{}\label{appenB}
\begin{lemma}\label{lemma_6}
\textit{(Sufficient Conditions for Type of Convergence \cite{galambos78, david03, sharif05, song06})} Let $\lambda_1,\lambda_2,\ldots,\lambda_K$
be i.i.d. random variables with CDF $F_{\lambda}(x)$. We denote $\Lambda_K=\max_i \lambda_i$. If there exists some distribution function $G$
which is nondegenerate and some constant $a_K\in \mathbb{R}, b_K>0$ such that the distribution of $\frac{\Lambda_K-a_K}{b_K}$ converges to $G$,
then $G$ must be one of just three types: $G_1$: Fr{\'e}chet distribution; $G_2$: Weibull distribution; $G_3$: Gumbel distribution.

The CDF of $\lambda_i$, i.e., $F_{\lambda}$ determines one of the exact types. If $F_{\lambda}$ results in one limiting distribution, then we
say $F_{\lambda}$ belongs to the domain of attraction of this type, i.e., $F_{\lambda}\in \mathcal{D}(G_i)$. The well-known sufficient
conditions for $F_{\lambda}\in \mathcal{D}(G_1)$ and $F_{\lambda}\in \mathcal{D}(G_3)$ are as follows: Define
$\omega(F_{\lambda})=\sup\{x:F_{\lambda}(x)<1\}$. $F_{\lambda}(x)$ is absolutely continuous and $f_{\lambda}(x)=F_{\lambda}'(x)$ and
$f_{\lambda}'(x)=F_{\lambda}''(x)$ exist, then

(a) $F_{\lambda}\in \mathcal{D}(G_1)$ if $f_{\lambda}(x)>0$ for all large $x$ and for some $\phi>0$,
\begin{equation}
\label{appen:eq_8} \lim_{x\rightarrow\infty}\frac{xf_{\lambda}(x)}{1-F_{\lambda}(x)}=\phi.
\end{equation}

(b) $F_{\lambda}\in \mathcal{D}(G_2)$ if $\mu<\infty$ and for some $\phi>0$,
\begin{equation}
\label{appen:eq_9} \lim_{x\rightarrow \mu}\frac{(\mu-x)f_{\lambda}(x)}{1-F_{\lambda}(x)}=\phi.
\end{equation}

(c) $F_{\lambda}\in \mathcal{D}(G_3)$ if $f_{\lambda}(x)>0$ and is differentiable for all $x$ in $(x_1,\omega(F_{\lambda}))$ for some $x_1$, and
\begin{equation}
\label{appen:eq_10} \lim_{x\rightarrow\omega(F_{\lambda})}\frac{d}{dx}\left[\frac{1-F_{\lambda}(x)}{f_{\lambda}(x)}\right]=0.
\end{equation}
Further, we can choose the normalizing constants $a_K=F_{\lambda}^{-1}(1-\frac{1}{K})$ and
$b_K=F_{\lambda}^{-1}(1-\frac{1}{Ke})-F_{\lambda}^{-1}(1-\frac{1}{K})$, where $F_{\lambda}^{-1}(x)=\inf\{y:F_{\lambda}(y)\geq x\}$.
\end{lemma}

\medskip

\textit{Proof of Theorem \ref{theorem_2}:} Assume $Z_k^{(0)}$ is a nonnegative random variable with CDF $F_{Z_k^{(0)}}(x)$ such that
$f_{Z_k^{(0)}}(x)>0$ and $f_{Z_k^{(0)}}'(x)$ exist. The random variable $Y_{k,M}^{(0)}$ is related to $Z_k^{(0)}$ by the following equation:
$F_{Y_{k,M}^{(0)}}(x)=\sum_{m=0}^{M-1}\xi_1(N,M,m)(F_{Z_k^{(0)}}(x))^{N-m}$. In order to show that $F_{Y_{k,M}^{(0)}}$ has the same type of
convergence property as $F_{Z_k^{(0)}}$, the proof in the sequel will be conducted for each of the three types.

(i) If for some $\phi>0$, $\mathop{\lim}\limits_{x\rightarrow\infty}\frac{xf_{Z_k^{(0)}}(x)}{1-F_{Z_k^{(0)}}(x)}=\phi$, then
$F_{Z_k^{(0)}}\in\mathcal{D}(G_1)$. It must be shown that
$\mathop{\lim}\limits_{x\rightarrow\infty}\frac{xf_{Y_{k,M}^{(0)}}(x)}{1-F_{Y_{k,M}^{(0)}}(x)}=\tilde{\phi}$ for some $\tilde{\phi}>0$.
Substituting the expression for $F_{Y_{k,M}^{(0)}}$ and $f_{Y_{k,M}^{(0)}}$ yields
\begin{align}
\mathop{\lim}\limits_{x\rightarrow\infty}\frac{xf_{Y_{k,M}^{(0)}}(x)}{1-F_{Y_{k,M}^{(0)}}(x)}&=\mathop{\lim}\limits_{x\rightarrow\infty}\frac{x\sum_{m=0}^{M-1}\xi_1(N,M,m)(N-m)(F_{Z_k^{(0)}}(x))^{N-m-1}f_{Z_k^{(0)}}(x)}{1-\sum_{m=0}^{M-1}\xi_1(N,M,m)(F_{Z_k^{(0)}}(x))^{N-m}}\notag\\
&\mathop{=}\limits^{(a)}\mathop{\lim}\limits_{x\rightarrow\infty}\frac{\sum_{m=0}^{M-1}\xi_1(N,M,m)(N-m)\left(f_{Z_k^{(0)}}(x)+xf_{Z_k^{(0)}}'(x)\right)}{-\sum_{m=0}^{M-1}\xi_1(N,M,m)(N-m)(F_{Z_k^{(0)}}(x))^{N-m-1}f_{Z_k^{(0)}}(x)}\notag\\
\label{appen:eq_11}&\mathop{=}\limits^{(b)}\frac{N-\sum_{m=0}^{M-1}\xi_1(N,M,m)m}{N-\sum_{m=0}^{M-1}\xi_1(N,M,m)m}\phi=\phi,
\end{align}
where (a) holds by applying the L'Hospital's rule; (b) follows from the type of convergence of $Z_k^{(0)}$. Therefore, $\tilde{\phi}=\phi$, and
$F_{Y_{k,M}^{(0)}}\in\mathcal{D}(G_1)$.

(ii) If $\mu<\infty$ and for some $\phi>0$, $\mathop{\lim}\limits_{x\rightarrow \mu}\frac{(\mu-x)f_{Z_k^{(0)}}(x)}{1-F_{Z_k^{(0)}}(x)}=\phi$,
then $F_{Z_k^{(0)}}\in\mathcal{D}(G_2)$. It must be shown that $\mathop{\lim}\limits_{x\rightarrow
\mu}\frac{(\mu-x)f_{Y_{k,M}^{(0)}}(x)}{1-F_{Y_{k,M}^{(0)}}(x)}=\tilde{\phi}$ for some $\tilde{\phi}>0$. Substituting the expression for
$F_{Y_{k,M}^{(0)}}$ and $f_{Y_{k,M}^{(0)}}$ yields
\begin{align}
\mathop{\lim}\limits_{x\rightarrow\mu}\frac{(\mu-x)f_{Y_{k,M}^{(0)}}(x)}{1-F_{Y_{k,M}^{(0)}}(x)}&=\mathop{\lim}\limits_{x\rightarrow\mu}\frac{(\mu-x)\sum_{m=0}^{M-1}\xi_1(N,M,m)(N-m)(F_{Z_k^{(0)}}(x))^{N-m-1}f_{Z_k^{(0)}}(x)}{1-\sum_{m=0}^{M-1}\xi_1(N,M,m)(F_{Z_k^{(0)}}(x))^{N-m}}\notag\\
&\mathop{=}\limits^{(a)}\mathop{\lim}\limits_{x\rightarrow\mu}\frac{\sum_{m=0}^{M-1}\xi_1(N,M,m)(N-m)(F_{Z_k^{(0)}}(x))^{N-m-1}\left(f_{Z_k^{(0)}}(x)-(\mu-x)f_{Z_k^{(0)}}'(x)\right)}{\sum_{m=0}^{M-1}\xi_1(N,M,m)(N-m)(F_{Z_k^{(0)}}(x))^{N-m-1}f_{Z_k^{(0)}}(x)}\notag\\
\label{appen:eq_12}&\mathop{=}\limits^{(b)}\mathop{\lim}\limits_{x\rightarrow\mu}\frac{\sum_{m=0}^{M-1}\xi_1(N,M,m)(N-m)(F_{Z_k^{(0)}}(x))^{N-m-1}}{\sum_{m=0}^{M-1}\xi_1(N,M,m)(N-m)(F_{Z_k^{(0)}}(x))^{N-m-1}}\phi=\phi,
\end{align}
where (a) holds by considering the term that dominant the limit and applying the L'Hospital's rule; (b) follows from the type of convergence of
$Z_k^{(0)}$. Therefore, $\tilde{\phi}=\phi$, and $F_{Y_{k,M}^{(0)}}\in\mathcal{D}(G_2)$.

(iii) If $\mathop{\lim}\limits_{x\rightarrow\infty}\frac{d}{dx}\left[\frac{1-F_{Z_k^{(0)}}(x)}{f_{Z_k^{(0)}}(x)}\right]=0$, and
$\mathop{\lim}\limits_{x\rightarrow\infty}\frac{1-F_{Z_k^{(0)}}(x)}{f_{Z_k^{(0)}}(x)}$ exists, then $F_{Z_k^{(0)}}\in\mathcal{D}(G_3)$. It must
be shown that $\mathop{\lim}\limits_{x\rightarrow\infty}\frac{d}{dx}\left[\frac{1-F_{Y_{k,M}^{(0)}}(x)}{f_{Y_{k,M}^{(0)}}(x)}\right]=0$.
Carrying out the differentiation, another equivalent condition is the following:
$\mathop{\lim}\limits_{x\rightarrow\infty}\frac{\left(F_{Y_{k,M}^{(0)}}(x)-1\right)f_{Y_{k,M}^{(0)}}'(x)}{(f_{Y_{k,M}^{(0)}}(x))^2}=1$. Substituting the
expression for $F_{Y_{k,M}^{(0)}}$ and $f_{Y_{k,M}^{(0)}}$ yields
\begin{equation}\label{appen:eq_13}
\begin{array}{ll}
&\mathop{\lim}\limits_{x\rightarrow\infty}\frac{\left(F_{Y_{k,M}^{(0)}}(x)-1\right)f_{Y_{k,M}^{(0)}}'(x)}{(f_{Y_{k,M}^{(0)}}(x))^2}\\
&=\mathop{\lim}\limits_{x\rightarrow\infty}\frac{\left(\sum_{m=0}^{M-1}\xi_1(N,M,m)(F_{Z_k^{(0)}}(x))^{N-m}-1\right)\left(\sum_{m=0}^{M-1}\xi_1(N,M,m)(N-m)(N-m-1)(F_{Z_k^{(0)}}(x))^{N-m-2}(f_{Z_k^{(0)}}(x))^2\right)}{\left(\sum_{m=0}^{M-1}\xi_1(N,M,m)(N-m)(F_{Z_k^{(0)}}(x))^{N-m-1}f_{Z_k^{(0)}}(x)\right)^2}\\
&\quad + \mathop{\lim}\limits_{x\rightarrow\infty}\frac{\left(\sum_{m=0}^{M-1}\xi_1(N,M,m)(F_{Z_k^{(0)}}(x))^{N-m}-1\right)\left(\sum_{m=0}^{M-1}\xi_1(N,M,m)(N-m)(F_{Z_k^{(0)}}(x))^{N-m-1}f_{Z_k^{(0)}}'(x)\right)}{\left(\sum_{m=0}^{M-1}\xi_1(N,M,m)(N-m)(F_{Z_k^{(0)}}(x))^{N-m-1}f_{Z_k^{(0)}}(x)\right)^2}\\
&\mathop{=}\limits^{(a)}\frac{\left(\sum_{m=0}^{M-1}\xi_1(N,M,m)(N-m)\right)^2}{\left(\sum_{m=0}^{M-1}\xi_1(N,M,m)(N-m)\right)^2}\frac{\left(\sum_{m=0}^{M-1}\xi_1(N,M,m)(F_{Z_k^{(0)}}(x))^{N-m}-1\right)\left(\sum_{m=0}^{M-1}\xi_1(N,M,m)(N-m)(F_{Z_k^{(0)}}(x))^{N-m-1}f_{Z_k^{(0)}}'(x)\right)}{(f_{Z_k^{(0)}}(x))^2}\\
&\mathop{=}\limits^{(b)}1,\\
\end{array}
\end{equation}
where (a) comes from the fact that $\sum_{m=0}^{M-1}\xi_1(N,M,m)=1$, (b) holds by applying the
L'Hospital's rule and the type of convergence of $Z_k^{(0)}$. Therefore, $F_{Y_{k,M}^{(0)}}\in\mathcal{D}(G_3)$.

\medskip

\textit{Proof of Corollary \ref{corollary_1}:} For the general $\mathsf{SINR}$ case, in order to prove that $F_{Z_k^{(0)}}\in\mathcal{D}(G_3)$,
it must be shown that $\mathop{\lim}\limits_{x\rightarrow\infty}\frac{d}{dx}\left[\frac{1-F_{Z_k^{(0)}}(x)}{f_{Z_k^{(0)}}(x)}\right]=0$.
Substituting the expression for $f_{Z_k^{(0)}}$ and $F_{Z_k^{(0)}}$ in (\ref{exact_eq_1}) and (\ref{exact_eq_2}) yields
\begin{align}
\mathop{\lim}\limits_{x\rightarrow\infty}\frac{1-F_{Z_k^{(0)}}(x)}{f_{Z_k^{(0)}}(x)}&=\mathop{\lim}\limits_{x\rightarrow\infty}\frac{\sum_{b=1}^{J_k}\frac{\varpi_k^{(b)}e^{-\frac{x}{\rho_k^{(0)}}}\rho_k^{(0)}}{\rho_k^{(0)}+\rho_k^{(b)}x}}{\sum_{b=1}^{J_k}\varpi_k^{(b)}e^{-\frac{x}{\rho_k^{(0)}}}\left(\frac{1}{\rho_k^{(0)}+\rho_k^{(b)}x}+\frac{\rho_k^{(0)}\rho_k^{(b)}}{\left(\rho_k^{(0)}+\rho_k^{(b)}x\right)^2}\right)}\notag\\
\label{appen:eq_14}&\mathop{=}\limits^{(a)}\frac{\rho_k^{(0)}\sum_{b=1}^{J_k}\varpi_k^{(b)}\frac{1}{\rho_k^{(b)}}}{\sum_{b=1}^{J_k}\varpi_k^{(b)}\frac{1}{\rho_k^{(b)}}}=\rho_k^{(0)},
\end{align}
where (a) follows from applying the L'Hospital's rule. It can be shown that
\begin{equation}\label{appen:eq_15}
f_{Z_k^{(0)}}'(x)=\sum_{b=1}^{J_k}\varpi_k^{(b)}e^{-\frac{x}{\rho_k^{(0)}}}\left(\frac{-\rho_k^{(0)}}{\rho_k^{(0)}+\rho_k^{(b)}x}+\frac{-2\rho_k^{(b)}}{\left(\rho_k^{(0)}+\rho_k^{(b)}x\right)^2}+\frac{-2\rho_k^{(0)}\rho_k^{(b)}}{\left(\rho_k^{(0)}+\rho_k^{(b)}x\right)^3}\right).
\end{equation}
Utilizing the same technique as in (\ref{appen:eq_14}), it can be easily shown that
\begin{equation}\label{appen:eq_16}
\mathop{\lim}\limits_{x\rightarrow\infty}\frac{f_{Z_k^{(0)}}'(x)}{f_{Z_k^{(0)}}(x)}=-\frac{1}{\rho_k^{(0)}}.
\end{equation}
By combining the results of (\ref{appen:eq_14}) and (\ref{appen:eq_16}), the following equation holds
$\mathop{\lim}\limits_{x\rightarrow\infty}\frac{\left(F_{Z_k^{(0)}}(x)-1\right)f_{Z_k^{(0)}}'(x)}{(f_{Z_k^{(0)}}(x))^2}=\frac{-\rho_k^{(0)}}{-\rho_k^{(0)}}=1$,
which proves the type of convergence of $F_{Z_k^{(0)}}$. Applying Theorem \ref{theorem_2} yields $F_{Y_{k,M}^{(0)}}\in\mathcal{D}(G_3)$.

\medskip

\textit{Proof of Corollary \ref{corollary_2}:} In the one-dominant interference limited case, it is easy to verify that
\begin{equation}\label{appen:eq_17}
\mathop{\lim}\limits_{x\rightarrow\infty}\frac{xf_{Z_k^{(0)}}(x)}{1-F_{Z_k^{(0)}}(x)}=\mathop{\lim}\limits_{x\rightarrow\infty}\frac{\rho_k^{(1)}x}{\rho_k^{(1)}x+\rho_k^{(0)}}=1.
\end{equation}
Thus, $F_{Z_k^{(0)}}\in\mathcal{D}(G_1)$. Applying Theorem \ref{theorem_2} yields $F_{Y_{k,M}^{(0)}}\in\mathcal{D}(G_1)$.

In the noise limited case, it is easy to verify that $F_{Z_k^{(0)}}\in\mathcal{D}(G_3)$, e.g., see \cite{song06}. Applying Theorem
\ref{theorem_2} yields $F_{Y_{k,M}^{(0)}}\in\mathcal{D}(G_3)$.

\medskip

\textit{Proof of Lemma \ref{lemma_5}:} To prove this lemma, the following theorem which discusses the extremes over random sample size is called
upon.
\begin{theorem} \label{theorem_4}
(\textit{Random Observations Theorem \cite{galambos78, berman62})} Let, as $K\rightarrow\infty$, $\frac{\kappa(K)}{K}\rightarrow\vartheta$ in
probability, where $\vartheta$ is a positive random variable. Assume that there are sequences $a_K\in \mathbb{R}, b_K>0$ such that
$\frac{\Lambda_K-a_K}{b_K}$ converges weakly to a nondegenerate distribution function $G$. Then, as $K\rightarrow\infty$,
\begin{equation}
\label{appen:eq_18} \lim\mathbb{P}\left(\Lambda_{\kappa(K)}<a_K+b_K x\right)=\int_{-\infty}^{\infty}G^{y}(x)d\mathbb{P}(\vartheta<y).
\end{equation}
\end{theorem}
From the analysis in Section \ref{rate}, when $K\rightarrow\infty$, $\frac{\kappa(K)}{K}\rightarrow\frac{M}{N}$. Thus from the above random
observations theorem, the extreme order statistics of the received CQI for a given user $k$ can be efficiently approximated by
$\left(F_{Y_{k,M}^{(0)}}\right)^{\frac{KM}{N}}$.

\medskip

\textit{Proof of Theorem \ref{theorem_3}:} According to the condition of domain of attraction, it must be shown that
\begin{equation}
\label{appen:eq_19} \lim_{r\rightarrow\infty}\frac{d}{dr}\left[\frac{1-F_{R_{k,M}}(r)}{f_{R_{k,M}}(r)}\right]=0.
\end{equation}
if $\mathop{\lim}\limits_{x\rightarrow\infty}\frac{xf_{Y_{k,M}}(x)}{1-F_{Y_{k,M}}(x)}=\phi>0$ or
$\mathop{\lim}\limits_{x\rightarrow\infty}\frac{d}{dx}\left[\frac{1-F_{Y_{k,M}}(x)}{f_{Y_{k,M}}(x)}\right]=0$.

It is derived in \cite{song06} that
\begin{align}
&\mathop{\lim}\limits_{r\rightarrow\infty}\frac{d}{dr}\left[\frac{1-F_{R_{k,M}}(r)}{f_{R_{k,M}}(r)}\right]\notag\\
&=\mathop{\lim}\limits_{r\rightarrow\infty}\left(-1-\frac{(1-F_{Y_{k,M}}(T^{-1}(r)))f_{Y_{k,M}}'(T^{-1}(r))}{(f_{Y_{k,M}}(T^{-1}(r)))^2}\right)-\mathop{\lim}\limits_{r\rightarrow\infty}\left(\frac{(1-F_{Y_{k,M}}(T^{-1}(r)))(T^{-1})''(r))}{f_{Y_{k,M}}(T^{-1}(r))((T^{-1})'(r))^2}\right)\notag\\
\label{appen:eq_20}&=\mathop{\lim}\limits_{x\rightarrow\infty}\frac{d}{dx}\left[\frac{1-F_{Y_{k,M}}(x)}{f_{Y_{k,M}}(x)}\right]-\mathop{\lim}\limits_{x\rightarrow\infty}\frac{1-F_{Y_{k,M}}(x)}{xf_{Y_{k,M}}(x)}.
\end{align}

If $\mathop{\lim}\limits_{x\rightarrow\infty}\frac{1-F_{Y_{k,M}}(x)}{xf_{Y_{k,M}}(x)}=\frac{1}{\phi}$, then $F_{Y_{k,M}}\in\mathcal{D}(G_1)$.
Using L'Hospital's rule, for a function $\theta(x)$ such as $\theta(x)\rightarrow\infty$ as $x\rightarrow\infty$, if
$\mathop{\lim}\limits_{x\rightarrow\infty}\frac{\theta(x)}{x}=\frac{1}{\phi}$, then
$\mathop{\lim}\limits_{x\rightarrow\infty}\theta'(x)=\frac{1}{\phi}$. This leads to
$\mathop{\lim}\limits_{r\rightarrow\infty}\frac{d}{dr}\left[\frac{1-F_{R_{k,M}}(r)}{f_{R_{k,M}}(r)}\right]=0$.

If $\mathop{\lim}\limits_{x\rightarrow\infty}\frac{d}{dx}\left[\frac{1-F_{Y_{k,M}}(x)}{f_{Y_{k,M}}(x)}\right]=0$, then
$F_{Y_{k,M}}\in\mathcal{D}(G_3)$. Similarly, applying L'Hospital's rule yields
$\mathop{\lim}\limits_{r\rightarrow\infty}\frac{d}{dr}\left[\frac{1-F_{R_{k,M}}(r)}{f_{R_{k,M}}(r)}\right]=0$.

Up to now, the sufficient conditions have been proved. From the analysis in Section \ref{rate} on the random effect of multiuser diversity due
to partial feedback, the number of CQI values fed back for each resource block becomes $\frac{KM}{N}$ with high probability. Additionally, note
that
\begin{equation}
\label{appen:eq_21} F_{R_{k,M}}^{-1}(x)=T(F_{Y_{k,M}}^{-1}(x))=\log_2\left(1+F_{Y_{k,M}}^{-1}(x)\right),
\end{equation}
thus the normalizing constants (\ref{asymptotic_eq_3}) can be obtained.

\medskip

\textit{Proof of Corollary \ref{corollary_3}:} In the one-dominant interference limited case with full feedback, the normalizing constants can
be obtained by using $F_{Y_{k,N}^{(0)}}^{-1}(x)=\frac{\rho_k^{(0)}x}{\rho_k^{(1)}(1-x)}$.

In the noise limited case with full feedback, the normalizing constants can be obtained by using
$F_{Y_{k,N}^{(0)}}^{-1}(x)=\rho_k^{(0)}\ln\frac{1}{1-x}$.

\medskip

\textit{Proof of Corollary \ref{corollary_4}:} In the one-dominant interference limited case with full feedback, the normalizing constants can
be obtained by using $F_{Y_{k,N}^{(0)}}^{-1}(x)=\frac{\rho_k^{(0)}x^{\frac{1}{N}}}{\rho_k^{(1)}(1-x^{\frac{1}{N}})}$ and the number of CQI
values fed back per resource block equaling $\frac{K}{N}$.

In the noise limited case with full feedback, the normalizing constants can be obtained by using
$F_{Y_{k,N}^{(0)}}^{-1}(x)=\rho_k^{(0)}\ln\frac{1}{1-x^{\frac{1}{N}}}$ and the number of CQI values fed back per resource block equaling
$\frac{K}{N}$.

\medskip

\textit{Proof of Corollary \ref{corollary_5}:} When $K\rightarrow\infty$, the probability of scheduling outage $(1-\frac{1}{N})^K\rightarrow 0$.
Therefore, from (\ref{asymptotic_eq_3}) it should be shown that
$\mathop{\lim}\limits_{K\rightarrow\infty}\frac{1-\frac{1}{K}}{(1-\frac{N}{K})^{\frac{1}{N}}}\rightarrow 1$. By applying the L'Hospital's rule,
the following equivalent equation holds:
\begin{equation}
\label{appen:eq_22} \mathop{\lim}\limits_{K\rightarrow\infty}\frac{\frac{1}{K}}{1-(1-\frac{N}{K})^{\frac{1}{N}}}=1.
\end{equation}

\medskip




\ifCLASSOPTIONcaptionsoff
  \newpage
\fi



%


%

\begin{IEEEbiography}[{\includegraphics[width=1in,height=1.25in,clip,keepaspectratio]{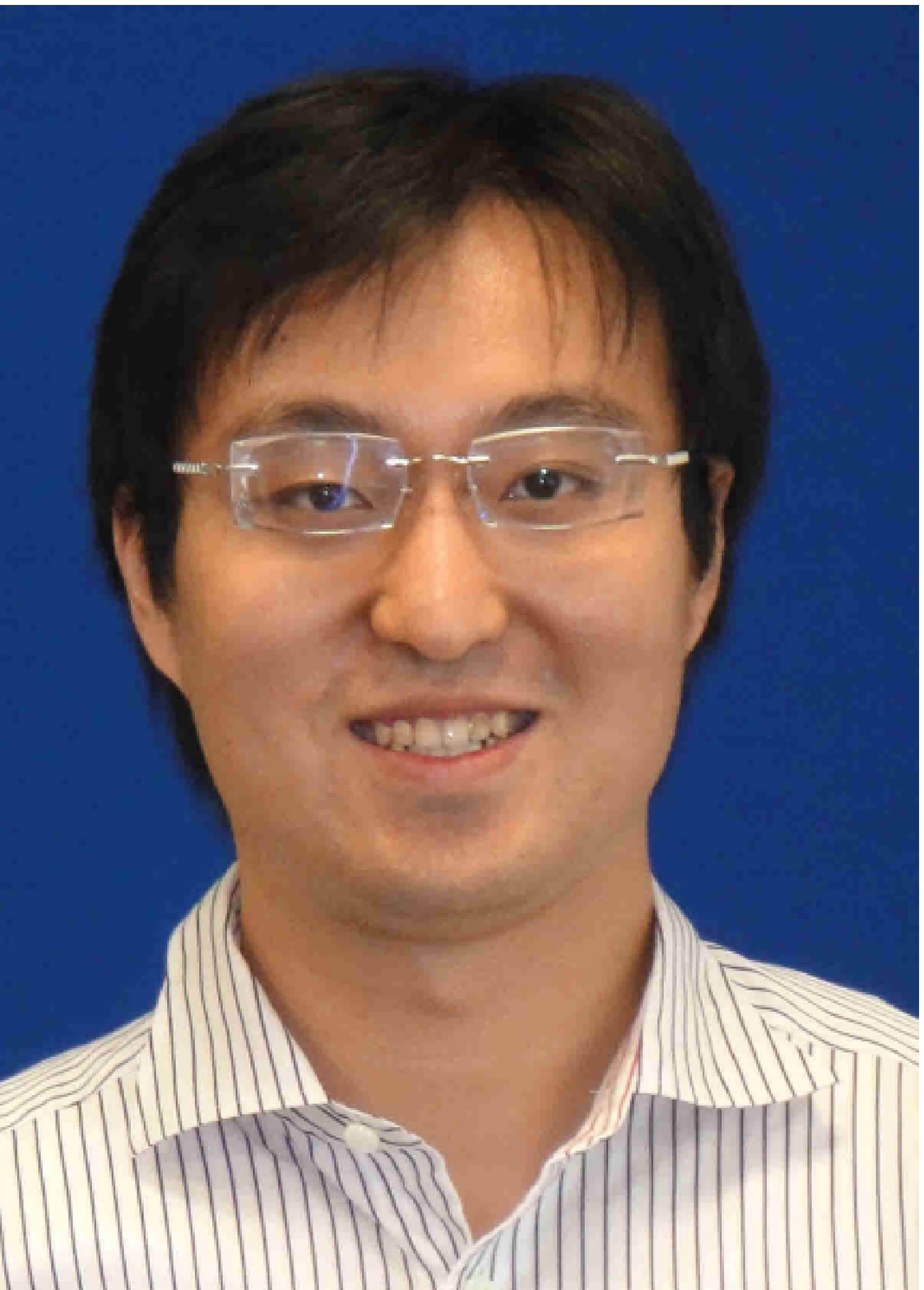}}]{Yichao Huang}
(S'10--M'12) received the B.Eng. degree in information engineering with highest honors from the Southeast University, Nanjing, China, in 2008,
and the M.S. and Ph.D. degrees in electrical engineering from the University of California, San Diego, La Jolla, in 2010 and 2012, respectively.

He interned with Qualcomm, Corporate R\&D, San Diego, CA, during summers 2011 and 2012. He was with California Institute for Telecommunications
and Information Technology (Calit2), San Diego, CA, during summer 2010. He was a visiting student at the Princeton University, Princeton, NJ,
during spring 2012. His research interests include communication theory, optimization theory, wireless networks, and signal processing for
communication systems.

Mr. Huang received the Presidential Scholarship both in 2005 and 2006, and the Best Thesis Award in 2008 from the Southeast University. He
received the Microsoft Young Fellow Award in 2007 from Microsoft Research Asia and the ECE Department Fellowship from the University of
California, San Diego in 2008 and was a finalist of Qualcomm Innovation Fellowship in 2010.
\end{IEEEbiography}
\begin{IEEEbiography}[{\includegraphics[width=1in,height=1.25in,clip,keepaspectratio]{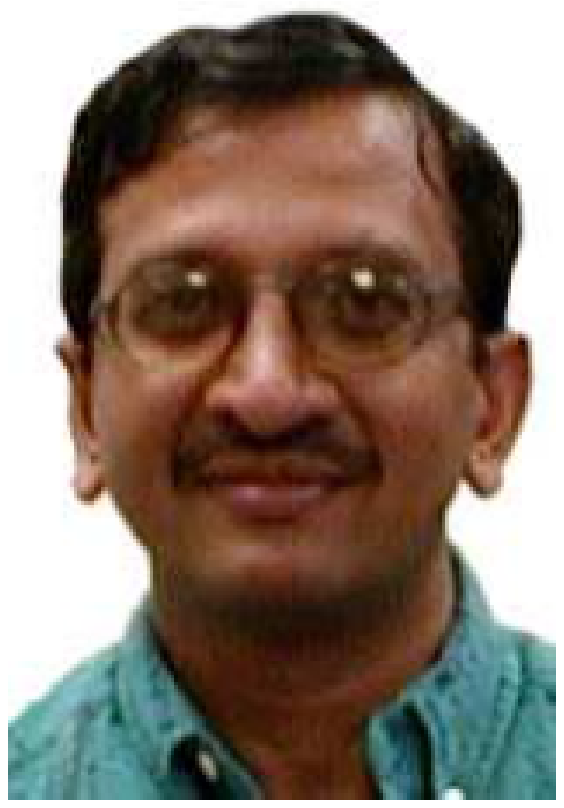}}]{Bhaskar D. Rao}
(S'80--M'83--SM'91--F'00) received the B.Tech. degree in electronics and electrical communication engineering from the Indian Institute of
Technology, Kharagpur, India, in 1979, and the M.Sc. and Ph.D. degrees from the University of Southern California, Los Angeles, in 1981 and
1983, respectively.

Since 1983, he has been with the University of California at San Diego, La Jolla, where he is currently a Professor with the Electrical and
Computer Engineering Department. He is the holder of the Ericsson endowed chair in Wireless Access Networks and was the Director of the Center
for Wireless Communications (2008--2011). His research interests include digital signal processing, estimation theory, and optimization theory,
with applications to digital communications, speech signal processing, and human--computer interactions.

Dr. Rao's research group has received several paper awards. His paper received the Best Paper Award at the 2000 Speech Coding Workshop and his
students have received student paper awards at both the 2005 and 2006 International Conference on Acoustics, Speech, and Signal Processing, as
well as the Best Student Paper Award at NIPS 2006. A paper he coauthored with B. Song and R. Cruz received the 2008 Stephen O. Rice Prize Paper
Award in the Field of Communications Systems. He was elected to the Fellow grade in 2000 for his contributions in high resolution spectral
estimation. He has been a Member of the Statistical Signal and Array Processing technical committee, the Signal Processing Theory and Methods
technical committee, and the Communications technical committee of the IEEE Signal Processing Society. He has also served on the editorial board
of the EURASIP Signal Processing Journal.
\end{IEEEbiography}







\end{document}